 \documentclass[epj]{svjour}

\newcommand\epjc[3]  {{Eur.\ Phys.\ J. }{\bf C #1} (#2) #3}

\newcommand\ijmpa[3] {{Int.\ J.\ Mod.\ Phys.\ }{\bf A #1} (#2) #3}

\newcommand\ncl[3]   {{Lett.\ Nuovo Cim.\ }{\bf #1} (#2) #3}

\newcommand\npa[3]   {{Nucl.\ Phys.\ }{\bf A #1} (#2) #3}
\newcommand\npb[3]   {{Nucl.\ Phys.\ }{\bf B #1} (#2) #3}

\newcommand\nc[3]    {{Nuovo Cim.\ }{\bf #1} (#2) #3}

\newcommand\plb[3]   {{Phys.\ Lett.\ }{\bf B #1} (#2) #3}

\newcommand\prd[3]   {{Phys.\ Rev.\ }{\bf D #1} (#2) #3}

\newcommand\prep[3]  {{Phys.\ Rept.\ }{\bf #1} (#2) #3}
\newcommand\prl[3]   {{Phys.\ Rev.\ Lett.\ }{\bf #1} (#2) #3}

\newcommand\ptp[3]   {{Prog.\ Theor.\ Phys.\ }{\bf #1} (#2) #3}

\newcommand\zpc[3]   {{Z.\ Physik }{\bf C #1} (#2) #3}

\newcommand{\hepph}[1]{{hep-ph/#1}}

\newcommand{\hepex}[1]{{hep-ex/#1}}

 \usepackage{epsfig}

 \title{
  Flavour structure of low-energy hadron pair photoproduction
  }

 \author{K.~Odagiri\inst{1}
  \and
  R.C.~Verma\inst{2}\fnmsep\thanks{Permanent address:
  Department of Physics, Punjabi University, Patiala-147002, India}
  }

 \institute{
  Institute of Physics, Academia Sinica, Nankang, Taipei, Taiwan 11529,
  The Republic of China
  \and
  Physics Division, National Center for Theoretical Sciences,
  Hsinchu, Taiwan 300, The Republic of China}

\begin{document}

 \abstract{
  We consider the process $\gamma\gamma\to H_1\overline H_2$ where $H_1$ 
and $H_2$ are either mesons or baryons.
  The experimental findings for such quantities as the $p\bar p$ and 
$K_SK_S$ differential cross sections, in the energy range currently 
probed, are found often to be in disparity with the scaling behaviour 
expected from hard constituent scattering.
  We discuss the long-distance pole--resonance contribution in 
understanding the origin of these phenomena, as well as the amplitude 
relations governing the short-distance contribution which we model as a 
scaling contribution.
  When considering the latter, we argue that the difference found for 
the $K_SK_S$ and the $K^+K^-$ integrated cross sections can be 
attributed to the $s$-channel isovector component.
  This corresponds to the $\rho\omega\to a$ subprocess in the VMD 
(vector-meson-dominance) language.
  The ratio of the two cross sections is enhanced by the suppression of
the $\phi$ component, and is hence constrained.
  We give similar constraints to a number of other hadron pair 
production channels.
  After writing down the scaling and pole--resonance contributions 
accordingly, the direct summation of the two contributions is found to 
reproduce some salient features of the $p\bar p$ and $K^+K^-$ data.
 %
 }

 \date{\today}


 \maketitle

 \section{Introduction}

  We consider the exclusive pair production process $\gamma\gamma\to
H_1\overline H_2$ where $H_1$ and $H_2$ are either mesons or baryons. We
consider the energy region not too far from the threshold, for example
the centre-of-mass energy $W_{\gamma\gamma}<4$~GeV.

  It has been found that some of the results of recent large-statistics 
measurement of these processes, for example at Belle 
\cite{bellep,bellek}, are difficult to reconcile with theoretical 
thinking based on hard constituent scattering. For example, there is a 
violation of the scaling behaviour expected from the naive 
quark-counting rule \cite{qcount},
 \begin{equation}
  \frac{d\sigma}{dt}\propto\frac{\cos\theta^*}{s^{K-2}},
  \label{eqn_quark_counting}
 \end{equation}
  where $K$ is the number of `elementary' fields taking part in the 
interaction. For instance, $K=1+1+2+2=6$ for $\gamma\gamma\to\pi\pi$. 
$s$ and $t$ are the usual Mandelstam variables. After integration over a 
constant $\cos\theta^*$ interval where $\theta^*$ is the polar angle of 
scattering in the centre-of-mass frame, this yields:
 \begin{equation}
  \sigma\propto1/s^3\ \mathrm{(mesons)},\qquad
  1/s^5\ \mathrm{(baryons)}.
  \label{eqn_photon_quark_counting}
 \end{equation}

\begin{figure}[ht]{
 \centerline{\epsfig{file=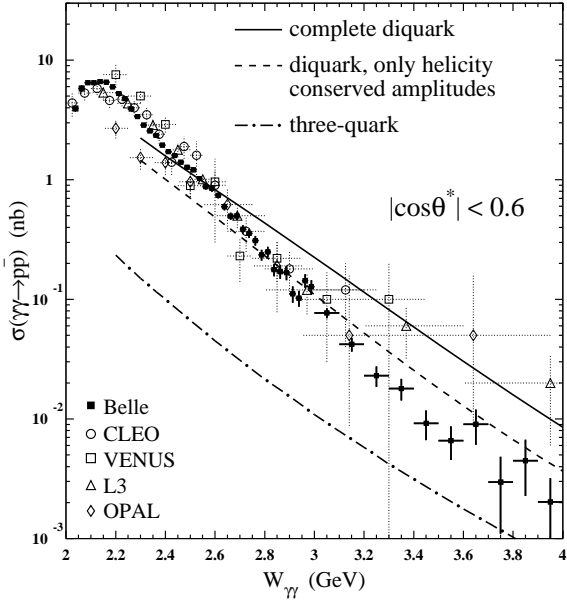,width=8cm}}
 \caption{$\gamma\gamma\to p\bar p$ cross section \cite{bellep} versus
centre-of-mass energy at VENUS \cite{venus}, CLEO \cite{cleo} and Belle
\cite{bellep} in the central region, defined by $|\cos\theta^*|<0.6$.
The vertical error-bars on the Belle data are due to the statistical
error in the event and the Monte Carlo samples only. Experimental data
is compared against three theoretical calculations, as described in
ref.~\cite{bellep}.
Figure reproduced with kind permission of the authors of 
ref.~\cite{bellep}.
 \label{fig_belle_crosssec}}}\end{figure}

  This scaling is expected to hold when the constituent-level hard 
subprocess approximately has a tree-level perturbative description. As 
seen in fig.~\ref{fig_belle_crosssec}, this fails for $\gamma\gamma\to 
p\bar p$ \cite{bellep,venus,cleo,l3ppbar} in the measured energy range.  
It seems to work for $\gamma\gamma\to K^+K^-$ \cite{bellek,tpc,argus} 
above $W_{\gamma\gamma}\approx2.4$~GeV but, disturbingly, seems to fail 
for $\gamma\gamma\to K_SK_S$ \cite{belleks} in the measured energy range 
of up to $W_{\gamma\gamma}\approx4$~GeV. The latter cross section drops 
faster with $W_{\gamma\gamma}$, and this finding, together with the 
large ratio between the two cross sections, is difficult to explain in 
frameworks based on hard constituent scattering. On the other hand, in 
the large-energy limit of the measured range, it has been found that the 
calculations of ref.~\cite{benayounchernyak}, based on leading-term QCD 
and wave functions following from the QCD sum rules, can accommodate the 
value found experimentally.

  Our main region of interest in this paper is below this energy range, 
where perturbative description is insufficient to account for the 
prominent features of the data. We are interested in the participation 
of alternative dynamics, that are more long-distance in nature, and are 
more appropriate to describing the observed distributions. At the same 
time, we are also interested in modelling the short-distance 
contribution with constraints from the amplitude factorization 
considerations.

  As a starting-point, let us consider VMD (vector-meson dominance). 
Here, the photon is interpreted as a quark-antiquark object, so that the 
exponent $K$ in eqn.~(\ref{eqn_quark_counting}) is modified. 
Corresponding to eqn.~(\ref{eqn_photon_quark_counting}), we would have:
 \begin{equation}
  \sigma\propto1/s^5\ \mathrm{(mesons)},\qquad
  1/s^7\ \mathrm{(baryons)}.
  \label{eqn_vmd_quark_counting}
 \end{equation}
  Although the applicability of the quark-counting rule to the VMD 
picture should not be taken for granted, this indicates that the fall in 
the cross section with the centre-of-mass energy would be more rapid 
than is expected from eqn.~(\ref{eqn_quark_counting}).
  It is a curious finding that for central events, defined by 
$|\cos\theta^*|<0.6$, the cross sections measured at Belle for 
$\gamma\gamma\to K_SK_S$ \cite{belleks} and $p\bar p$ \cite{bellep} go 
as $\sim W_{\gamma\gamma}^{-(8\sim12)}$ and $\sim 
W_{\gamma\gamma}^{-(12\sim15)}$ respectively for some regions of 
$W_{\gamma\gamma}$ away from the resonance region. As the exponent is 
sensitive to the cut on $|\cos\theta^*|$, it is possible that this 
agreement with eqn.~(\ref{eqn_vmd_quark_counting}) is accidental. We 
note nevertheless that a result of the form above can be derived from 
the consideration of the Sudakov form-factor effects 
\cite{brodskylepage,pinch}.

  Encouraged by this finding, we go on to consider the factorization of 
the scattering amplitude into the production and decay parts. The 
production subprocess is dominated, in case of ideal mixing between 
$\omega$ and $\phi$, by:
 \begin{eqnarray}
  \rho^0\rho^0,\omega\omega&\to&f_{ud}, \label{eqn_production_a} \\
  \rho^0\omega,\omega\rho^0&\to&a, \label{eqn_production_b} \\
  \phi\phi&\to&f_s. \label{eqn_production_c}
 \end{eqnarray}
  In the above, $f_{ud}$ stands for the $(u\bar u+d\bar d)/\sqrt{2}$ 
state. After relaxing the condition of ideal mixing, $f_{ud}$ and $f_s$ 
mix to give the physical $f$ and $f'$ mesons.
  We are not necessarily adopting the $s$-channel resonance picture, and 
$f$ and $a$ are, for now, merely a label of the $s$-channel flavour 
structure.

 \begin{figure}[ht]{
 \centerline{\epsfig{file=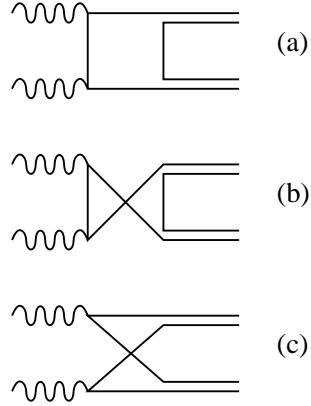,width=4cm}}
 \caption{The quark-line diagrams for meson-pair production. Similar 
diagrams can be drawn for the baryon-pair case. Diagrams (a) and (b) 
have $s$-channel representation whereas the 4-quark mode (c) does not.
 \label{fig_qline_meson}}}\end{figure}

  The quark-line diagrams are shown in fig.~\ref{fig_qline_meson}.
  The production-decay factorization assumption implicitly corresponds 
to the choice of `handbag' diagrams (a) and (b) of 
fig.~\ref{fig_qline_meson}, neglecting the 4-quark intermediate state of 
`cat's-ears' diagram (c).
  When considering long-distance dynamics, due to the degeneracy of the 
meson trajectories, we shall argue later that this approximation is 
acceptable for meson-pair production but not for baryon-pair production.
  On the other hand, while calculating the short-distance contribution, 
the relative size of the cat's-ears diagram is sensitive to the 
factorization scheme adopted.
  Here we limit ourselves to remarking that in the findings of 
ref.~\cite{handbag}, and for the energy range currently probed by 
experiment, their so-named `handbag' approach is appropriate for 
describing some features of the data at hand. For instance, the angular 
distribution for the $K_SK_S$ process, as seen in ref.~\cite{belleks}, 
is in better agreement than the more traditional approach of 
ref.~\cite{benayounchernyak}, although the overall normalization is not 
well understood in either of the two approaches.

  We have also neglected the OZI-suppressed channels such as
$\rho^0\phi\to f$.
  The decay part can be expressed similarly.

  We see immediately that any difference between the $K^+K^-$ and 
$K^0\overline{K^0}=K_LK_L+K_SK_S$ cross sections must be due to the 
simultaneous presence of the isoscalar $f/f'$ and isovector $a=(u\bar 
u-d\bar d)/\sqrt{2}$ components. It is also not difficult to see that 
the large ratio between $K^+K^-$ and $K_SK_S$ processes can be increased 
by the suppression of $\phi\phi\to f_s$.

  We proceed by modelling the short-distance piece as a scaling 
contribution obeying 
eqns.~(\ref{eqn_quark_counting},\ref{eqn_photon_quark_counting}).
  The long-distance piece includes the resonances of 
eqns.~(\ref{eqn_production_a})--(\ref{eqn_production_c}). These are 
related to the $t$-channel pole picture by duality, so that we can also 
model them as Regge amplitudes \cite{collins,ddln}. The signature term 
in the latter case would then represent the `cat's-ears' contributions.

  We find that the direct summation of the two amplitudes reproduces 
some of the salient features of the $K^+K^-$ and $p\bar p$ data, 
although the description of the angular distribution in the former case 
is poor.

  The modelling of the full amplitude as the sum of the two 
contributions at first sight may seem to suffer from the problem of 
double counting. However, we find numerically that adopting the 
alternative approach of a form factor that interpolates between the two 
amplitude in general cannot yield these results. An intuitive 
explanation would be that the finite-time short-distance effects and 
infinite-time Regge-pole dynamics have little overlap. 
  More explicitly, short-distance amplitudes involve a finite and small 
number of intermediate particles that individually carry large 
virtuality, and are independent from the long-distance amplitudes where 
virtuality is assigned, in the parton language, collectively to the 
intermediate state partons.
  Hence in our understanding, any discrepancy with the data that arise 
are due to our ignorance of low-energy dynamics, rather than being due 
to some form of double-counting in this summation procedure.
  In any case, such approach is not new. For a well-known example 
involving the summation of short-distance and long-distance amplitudes, 
see ref.~\cite{dl_tripleglu}.

  This paper is organized as follows. In sec.~\ref{sec_clebschgordan}, 
we present the relevant Clebsch-Gordan coefficients for kaons and for 
other mesons and baryons.
  In sec.~\ref{sec_dynamics}, we consider the effect of pole(-resonance) 
dynamics.
  The conclusions are stated at the end.

 \section{$SU(3)$ analysis} \label{sec_clebschgordan}

  We decompose the $(2\to2)$ amplitudes into the $s$-channel production
part and the decay part, so that, for example, the $s$-channel scalar
part of the amplitudes is given by:
 \begin{eqnarray}
  &&\mathcal A(\gamma\gamma\to V_1V_2\to S\to H_1\overline{H}_2)
  \nonumber\\&&\qquad\qquad=
  \gamma_{V_1}^{-1}\gamma_{V_2}^{-1}
  g_{V_1V_2S}\times g_{SH_1H_2}\times F(S).
  \label{eqn_amplitude_factorization}
 \end{eqnarray}
  $g$ are proportional to the $SU(3)$ flavour Clebsch-Gordan 
coefficients, whereas the dynamics is contained in the function $F(S)$.
  $\gamma_V$ are the photon-$V$ coupling constants satisfying:
 \begin{equation}
  \gamma_{\rho^0}^{-1}:\gamma_\omega^{-1}:\gamma_\phi^{-1}
  \approx 3:1:-\sqrt2.
  \label{eqn_VMD_couplings}
 \end{equation}
  We shall adopt the ratio $3:1$ for the $\rho^0$ and $\omega$ couplings 
later on, but modify the $\phi$ coupling by a suppression factor 
$\sqrt{\delta}$. It should be understood that this suppression factor is 
introduced in order to suppress $f'$ contribution. The photon coupling 
to $\phi$ is not changed.

  We note that the $p\bar p$ data \cite{bellep} shows clear indication 
of the presence of a pseudoscalar $\eta_c$ resonance, and the same peak 
is also present in the $K\overline K$ data \cite{bellek}. The 
pseudoscalar contribution can be included without modifying the 
structure of the formalism.

 \subsection{Production subprocess}

  $u,d$ and $s$ quarks are organized into a flavour triplet structure 
as:
 \begin{equation}
  \mathbf{3}\equiv q_a\equiv
  \left(\begin{array}{c} u \\ d \\ s\end{array}\right), \quad
  \mathbf{3^*}\equiv \bar q^a \equiv
  \left(\begin{array}{ccc} \bar u & \bar d & \bar s\end{array}\right).
 \end{equation}
  Mesons are in $\mathbf3\otimes\mathbf{3^*}$ and baryons are in
$\mathbf3\otimes\mathbf3\otimes\mathbf3$.

  We define the nonet $1^-,0^-$, and $0^+$ mesons by $V^a_b, P^a_b$, and
$S^a_b$, respectively. These are constructed explicitly as:
 \begin{eqnarray}
  &&\mathbf3\otimes\mathbf{3^*}\equiv
  \left(\begin{array}{ccc} u\bar u & u\bar d & u\bar s \\
  d\bar u & d\bar d & d\bar s \\ s\bar u & s\bar d & s\bar s
  \end{array}\right), \\
  &&V^a_b=
  \nonumber\\&&
  \left(\begin{array}{ccc}
  \frac{\rho^0+\omega\cos\theta_V-\phi\sin\theta_V}{\sqrt2}
  \hspace*{-0.5cm}
  & \rho^+ & K^{*+} \\
  \rho^- &
  \hspace*{-0.5cm}
  \frac{-\rho^0+\omega\cos\theta_V-\phi\sin\theta_V}{\sqrt2}
  \hspace*{-0.5cm}
  & K^{*0}  \\
  K^{*-} & \overline{K^{*0}} & \omega\sin\theta_V+\phi\cos\theta_V
  \end{array}\right),
  \label{eqn_vector_nonet} \\
  &&P^a_b=
  \nonumber\\&&
  \left(\begin{array}{ccc}
  \frac1{\sqrt2}\pi^0+\frac1{\sqrt6}\eta_8+\frac1{\sqrt3}\eta_1
  \hspace*{-0.5cm}
  & \pi^+ & K^+ \\  \pi^- &
  \hspace*{-0.5cm}
  -\frac1{\sqrt2}\pi^0+\frac1{\sqrt6}\eta_8+\frac1{\sqrt3}\eta_1
  \hspace*{-0.5cm}
  & K^0 \\
  K^- & \overline{K^0} & -\frac2{\sqrt6}\eta_8+\frac1{\sqrt3}\eta_1
  \end{array}\right),
  \label{eqn_pseudoscalar_nonet}\\
  &&S^a_b=\left(\begin{array}{ccc} u\bar u & a_0^+ & K_0^+ \\
  a_0^- & d\bar d & K_0^0 \\ K_0^- & \overline{K_0^0} &
  s\bar s \end{array}\right).
  \label{eqn_scalar_nonet}
 \end{eqnarray}
  We have defined the $\omega-\phi$ mixing angle $\theta_V$ in the 
above. For ideal mixing, $\theta_V=0$.

  The Clebsch-Gordan coefficients for the production subprocesses of
eqns.~(\ref{eqn_production_a})--(\ref{eqn_production_c}) are calculated
by the contribution of the diagonal, $a=b=c$, part of the quantity:
 \begin{equation}
  \frac12\left(V^c_bV^a_c+V^a_cV^c_b\right)S^b_a.
 \end{equation}
  This particular notation implies spin-0 state in the $s$-channel,
but the structure of the expression is general to any even-spin positive
parity states.
  The $f-f'$ mixing angle $\theta_S$ is defined by:
 \begin{eqnarray}
  a_0&=&(u\bar u-d\bar d)/\sqrt2,\\
  f_0&=&\cos\theta_S(u\bar u+d\bar d)/\sqrt2+\sin\theta_S(s\bar s),\\
  f_0'&=&-\sin\theta_S(u\bar u+d\bar d)/\sqrt2+\cos\theta_S(s\bar s).
 \end{eqnarray}
  For later use, we define $\theta_P$ analogously, which describes 
$\eta-\eta'$ mixing.

  The Clebsch-Gordan coefficients can now be calculated, and these are 
listed in tab.~\ref{tab_VVS}. We show the general case as well as the 
two cases of ideal mixing, corresponding to $\theta_V=0$ and to 
$\theta_V=\theta_S=0$. In reality, $\theta_V=0$ is a good approximation, 
although $\theta_S=0$ is doubtful for the low-lying resonances.
  The terms that vanish in the ideal mixing case are OZI-suppressed. 
 \begin{table*}[ht]{
  \begin{center}\begin{tabular}{cccc}
   \hline
   vertex & general mixing & $\theta_V=0$ & $\theta_V=\theta_S=0$ \\
   \hline
   $\rho^0\rho^0\to f_0$ &
   $\cos\theta_S/\sqrt2$ & $\cos\theta_S/\sqrt2$ & $1/\sqrt2$ \\
   $\rho^0\rho^0\to f_0'$ &
   $-\sin\theta_S/\sqrt2$ & $-\sin\theta_S/\sqrt2$ & $0$ \\
   $\rho^0\omega,\omega\rho^0\to a_0$ &
   $\cos\theta_V/\sqrt2$ & $1/\sqrt2$ & $1/\sqrt2$ \\
   $\rho^0\phi,\phi\rho^0\to a_0$ &
   $-\sin\theta_V/\sqrt2$ & $0$ & $0$ \\
   \hline
   $\omega\omega\to f_0$ &
   $\cos^2\theta_V\cos\theta_S/\sqrt2+\sin^2\theta_V\sin\theta_S$ &
   $\cos\theta_S/\sqrt2$ & $1/\sqrt2$ \\
   $\omega\omega\to f_0'$ &
   $-\cos^2\theta_V\sin\theta_S/\sqrt2+\sin^2\theta_V\cos\theta_S$ &
   $-\sin\theta_S/\sqrt2$ & $0$ \\
   $\omega\phi\to f_0$ &
   $\sin\theta_V\cos\theta_V(-\cos\theta_S/\sqrt2+\sin\theta_S)$ &
   $0$ & $0$ \\
   $\omega\phi\to f_0'$ &
   $\sin\theta_V\cos\theta_V(\sin\theta_S/\sqrt2+\cos\theta_S)$ &
   $0$ & $0$ \\
   \hline
   $\phi\phi\to f_0$ &
   $\sin^2\theta_V\cos\theta_S/\sqrt2+\cos^2\theta_V\sin\theta_S$ &
   $\sin\theta_S$ & $0$ \\
   $\phi\phi\to f_0'$ &
   $-\sin^2\theta_V\sin\theta_S/\sqrt2+\cos^2\theta_V\cos\theta_S$ &
   $\cos\theta_S$ & $1$ \\
   \hline
  \end{tabular}\end{center}
  \caption{The $VVS$ coupling coefficients. The overall symmetry factor
2 has been suppressed.
  The vertexes not listed here are forbidden by isospin conservation, so
that we have:
   $g_{\rho^0\rho^0 a_0}=g_{\rho^0\omega f_0}=g_{\rho^0\omega f_0'}=
   g_{\rho^0\phi f_0}=g_{\rho^0\phi f_0'}=g_{\omega\omega a_0}=
   g_{\omega\phi a_0}=g_{\phi\phi a_0}=0$. We list the general case as 
well as the `ideal mixing' case corresponding to $\theta_V=0$ and to
$\theta_V=\theta_S=0$.
  \label{tab_VVS}
  }
 }\end{table*}

  One consequence of the OZI suppression is that when we consider the 
full production process, namely $\gamma\gamma\to V_1V_2$ with the vector 
mesons in the final state, the channels which necessarily involve 
OZI-suppressed interaction are suppressed, so that we expect:
 \begin{eqnarray}
  &&\sigma(\gamma\gamma\to\rho^0\rho^0),
  \sigma(\gamma\gamma\to\omega\omega)
  \nonumber\\&&\qquad\qquad>
  \sigma(\gamma\gamma\to\rho^0\omega)>
  \sigma(\gamma\gamma\to\phi\phi)
  \nonumber\\&&\qquad\qquad\qquad\qquad>\!\!>
  \sigma(\gamma\gamma\to\rho^0\phi),
  \sigma(\gamma\gamma\to\omega\phi).
 \end{eqnarray}
  The reasoning goes as follows. Let us first emphasize that this is for 
the entire $s$-channel production process as given by the factorization 
of eqn.~(\ref{eqn_amplitude_factorization}). The photons first couple to 
the appropriate vector boson, which fuse together into a (scalar) 
resonance, then finally decay into the states given above. In the ideal 
mixing case, $\rho^0\rho^0$ and $\omega\omega$ both come from $f$ decay 
with equal strength, so that $\rho^0\rho^0$ and $\omega\omega$ are 
approximately equal. $\rho^0\omega$ can come from $a$, but $a$ 
production is slightly smaller than $f$ because of the ratio of $\rho^0$ 
and $\omega$ contents of the photon. $\phi\phi$ is suppressed because 
$f'$ is less abundant than $f$, that is, $\rho^0\rho^0\to f$ is the 
dominant resonance production subprocess.

 \subsection{Meson pair production}

  We now turn to the decay subprocess. Let us first consider the
production of pseudo-scalar mesons. These are in the nonet representation 
of eqn.~(\ref{eqn_pseudoscalar_nonet}).

  The relevant coefficients can be obtained by the diagonal, $a=b$, part
of:
 \begin{equation}
  \frac12S^b_a(P^c_bP^a_c+P^a_cP^c_b).
 \end{equation}
  We list them in tab.~\ref{tab_SPP}.
 \begin{table*}[ht]{
  \begin{center}\begin{tabular}{cccc}
   \hline
    final state  & $a_0$ & $f_0$ & $f_0'$ \\
   \hline
    $\pi^+\pi^-$ & $0$ & $\cos\theta_S\sqrt2$ & $-\sin\theta_S\sqrt2$ \\
    $K^+K^-$ & $\frac1{\sqrt2}$ &
     $\frac{\cos\theta_S}{\sqrt2}+\sin\theta_S$ &
     $-\frac{\sin\theta_S}{\sqrt2}+\cos\theta_S$ \\
    $\pi^0\pi^0$ & $0$ & $\cos\theta_S\sqrt2$ & $-\sin\theta_S\sqrt2$ \\
    $K^0\overline{K^0}$ & $-\frac1{\sqrt2}$ &
     $\frac{\cos\theta_S}{\sqrt2}+\sin\theta_S$ &
     $-\frac{\sin\theta_S}{\sqrt2}+\cos\theta_S$ \\
   \hline
    $\pi^0\eta,\eta\pi^0$ & $\cos\theta_P\sqrt2$ & $0$ & $0$ \\
    $\pi^0\eta',\eta'\pi^0$ & $-\sin\theta_P\sqrt2$ & $0$ & $0$ 
    \\
   \hline
    $\eta\eta$ & $0$ &
     $2\left(\cos^2\theta_P\frac{\cos\theta_S}{\sqrt2}
     +\sin^2\theta_P\sin\theta_S\right)$ &
    $2\left(-\cos^2\theta_P\frac{\sin\theta_S}{\sqrt2}
     +\sin^2\theta_P\cos\theta_S\right)$ \\
    $\eta\eta'$ & $0$ & $2\sin\theta_P\cos\theta_P
     (-\frac{\cos\theta_S}{\sqrt2}+\sin\theta_S)$ &
     $2\sin\theta_P\cos\theta_P
     (\frac{\sin\theta_S}{\sqrt2}+\cos\theta_S)$ \\
    $\eta'\eta'$ & $0$ &
     $2\left(\sin^2\theta_P\frac{\cos\theta_S}{\sqrt2}
     +\cos^2\theta_P\sin\theta_S\right)$ &
     $2\left(-\sin^2\theta_P\frac{\sin\theta_S}{\sqrt2}
     +\cos^2\theta_P\cos\theta_S\right)$ \\
   \hline
  \end{tabular}\end{center}
  \caption{The $SPP$ coupling coefficients.
  \label{tab_SPP}
  }
 }\end{table*}

  Again, those modes that are OZI suppressed are accompanied by factor
$\sin\theta_S$ or $\sin\theta_P$. However, since $\theta_P$ is now
considerably large, $\approx-39$~degrees \cite{thetap}, the suppression
factor is only moderate. $\theta_S$ is also considerably large for the
spin-$0$ bosons \cite{pdg} although possibly not for the higher-spin
excitations of $a$, $f$ and $f'$.

  We see that, after including the identical particle factor of $1/2$
for the $\pi^0\pi^0$ cross section:
 \begin{equation}
  \sigma(\gamma\gamma\to\pi^+\pi^-)=2\sigma(\gamma\gamma\to\pi^0\pi^0),
 \end{equation}
  regardless of the mixing angles or production dynamics. We also see 
that the channels:
 \begin{equation}
  \gamma\gamma\to \pi^0\eta,\pi^0\eta',
 \end{equation}
  can only proceed via $s$-channel isovector $a$, so that the 
observation of these processes would be interesting to confirm the 
presence of the isovector channel. The magnitude of these channels is 
related to the difference of the kaon amplitudes, i.e.:
 \begin{eqnarray}
  &&\left|\mathcal A(\gamma\gamma\to K^+K^-)-
  \mathcal A(\gamma\gamma\to K^0\overline{K^0})\right|^2
  \nonumber\\&&\qquad=
  \left|\mathcal A(\gamma\gamma\to\pi^0\eta)\right|^2+
  \left|\mathcal A(\gamma\gamma\to\pi^0\eta')\right|^2,
  \label{eqn_SPP_sumrule_a}
 \end{eqnarray}
  In particular, when the $K^+K^-$ cross section dominates over
$K^0\overline{K^0}$, and when $W_{\gamma\gamma}$ is sufficiently above 
the $\pi^0\eta'$ threshold,
  we obtain:
 \begin{equation}
  \sigma(\gamma\gamma\to K^+K^-)\approx
  \sigma(\gamma\gamma\to\pi^0\eta)+\sigma(\gamma\gamma\to\pi^0\eta').
 \end{equation}
  This should be tested experimentally. However, we shall show later 
that the ratio between the $K^+K^-$ and $K^0\overline{K^0}$ amplitudes 
is expected to be at most 4, and hence omitting the $K^0\overline{K^0}$ 
amplitude contribution is not a good approximation. One possible 
assumption would be that the two cross sections only differ by a real 
constant factor. The cross section would then be scaled by:
 \begin{eqnarray}
  &&\sigma(\gamma\gamma\to K^+K^-)\longrightarrow
   \nonumber\\&&\qquad
   \sigma(\gamma\gamma\to K^+K^-)\times
   \left(1-\sqrt{\frac{\sigma(\gamma\gamma\to K^0\overline{K^0})}
   {\sigma(\gamma\gamma\to K^+K^-)}}\right)^2.
   \label{eqn_crosssec_scaling}
 \end{eqnarray}
  Similarly, for the sum of the $K^+K^-$ and $K^0\overline{K^0}$ 
amplitudes, we find that:
 \begin{eqnarray}
  &&2\left[\mathcal A(\gamma\gamma\to K^+K^-)+
  \mathcal A(\gamma\gamma\to K^0\overline{K^0})\right]
  \nonumber\\&&\qquad=
  \mathcal A(\gamma\gamma\to \eta\eta)+
  \mathcal A(\gamma\gamma\to \eta'\eta').
  \label{eqn_SPP_sumrule_b}
 \end{eqnarray}
  However, this equality would be difficult to test experimentally, 
unless it is found, for instance, that $\eta\eta$ dominates over 
$\eta'\eta'$ when sufficiently above the threshold. In this case, the 
relation would reduce to:
 \begin{equation}
  \sigma(\gamma\gamma\to K^+K^-)\approx
  \frac12\sigma(\gamma\gamma\to\eta\eta).
 \end{equation}
  The $\frac12$ factor on the right-hand side comes from the combination 
of the factor $2$ in eqn.~(\ref{eqn_SPP_sumrule_b}) and the factor 
$\frac12$ for identical particle production.
  Again, we can adjust for the error in neglecting the sub-leading 
amplitudes by a scaling similar to eqn.~(\ref{eqn_crosssec_scaling}) but 
with the minus sign in the brackets replaced by a plus sign.

  The above argument holds regardless of the production subprocess
$VV\to S$. Let us now consider the inclusion of the production
subprocess. This is obtained by referring to
eqn.~(\ref{eqn_amplitude_factorization}). We adopt ideal mixing for the
vector mesons so that $\theta_V=0$. The result, for $\gamma\gamma\to
K^+K^-, K^0\overline{K^0}, \pi^+\pi^-$, and $\pi^0\pi^0$ are:
 \begin{eqnarray}
  &&\mathcal A(\gamma\gamma\to K^+K^-)
  \nonumber\\&&\qquad=
   F(f_0) \left(\frac{\cos\theta_S}{\sqrt2}+\sin\theta_S\right)
   \left(\frac59\frac{\cos\theta_S}{\sqrt2}+
   \frac\delta9\sin\theta_S\right)\nonumber\\&&\qquad+
   F(f_0') \left(\frac{\sin\theta_S}{\sqrt2}-\cos\theta_S\right)
   \left(\frac59\frac{\sin\theta_S}{\sqrt2}-
   \frac\delta9\cos\theta_S\right)\nonumber\\&&\qquad+\frac16F(a_0)
  ,\\ &&\mathcal A(\gamma\gamma\to K^0\overline{K^0})
  \nonumber\\&&\qquad=
   F(f_0) \left(\frac{\cos\theta_S}{\sqrt2}+\sin\theta_S\right)
   \left(\frac59\frac{\cos\theta_S}{\sqrt2}+
   \frac\delta9\sin\theta_S\right)\nonumber\\&&\qquad+
   F(f_0') \left(\frac{\sin\theta_S}{\sqrt2}-\cos\theta_S\right)
   \left(\frac59\frac{\sin\theta_S}{\sqrt2}-
   \frac\delta9\cos\theta_S\right)\nonumber\\&&\qquad-\frac16F(a_0)
  ,\\ &&\mathcal A(\gamma\gamma\to \pi^+\pi^-)
  \nonumber\\&&\qquad=
   F(f_0) \cos\theta_S\left(\frac59\cos\theta_S+
   \frac\delta9\sqrt2\sin\theta_S\right)\nonumber\\&&\qquad+
   F(f_0') \sin\theta_S\left(\frac59\sin\theta_S-
   \frac\delta9\sqrt2\cos\theta_S\right)
  ,\\ &&\mathcal A(\gamma\gamma\to \pi^0\pi^0)
  \nonumber\\&&\qquad=
   F(f_0) \cos\theta_S\left(\frac59\cos\theta_S+
   \frac\delta9\sqrt2\sin\theta_S\right)\nonumber\\&&\qquad+
   F(f_0') \sin\theta_S\left(\frac59\sin\theta_S-
   \frac\delta9\sqrt2\cos\theta_S\right)
  .
 \end{eqnarray}
  $\delta$ is defined below eqn.~(\ref{eqn_VMD_couplings}).
  The above expressions simplify in the case $\theta_S=0$. As mentioned 
earlier, this becomes acceptable for the higher-spin excitations of 
$f,a$, and $f'$. We then have:
 \begin{eqnarray}
  &&\mathcal A(\gamma\gamma\to K^+K^-)\nonumber\\&&\qquad\qquad=
   \frac1{18}\left[5 F(f_0) + 2\delta F(f_0') + 3 F(a_0)\right]
  ,\\ &&\mathcal A(\gamma\gamma\to K^0\overline{K^0})
  \nonumber\\&&\qquad\qquad=
   \frac1{18}\left[5 F(f_0) + 2\delta F(f_0') - 3 F(a_0)\right]
  ,\\ &&\mathcal A(\gamma\gamma\to \pi^+\pi^-)= \frac59F(f_0)
  ,\\ &&\mathcal A(\gamma\gamma\to \pi^0\pi^0)= \frac59F(f_0).
 \end{eqnarray}
  We may simplify further by the approximation $F(f_0)=F(a_0)$ and
absorbing the difference between $F(f_0')$ and $F(f_0)$ in the 
coefficient $\delta$.
  Under these conditions, i.e., $F(a)=F(f)=F(f')$ for arbitrary 
$\delta$, it turns out that all dependence on $\theta_S$ cancels 
automatically, so that ideal mixing becomes a redundant assumption.
  We obtain, at the amplitude level:
 \begin{equation}
  K^+K^- : K^0\overline{K^0} : \pi^+\pi^- : \pi^0\pi^0 =
  4+\delta : 1+\delta : 5 : 5.
  \label{eqn_PP_amp_ratios}
 \end{equation}
  In particular, for the ratio of the $\gamma\gamma\to K^+K^-$ and
$K_SK_S$ cross sections, we have:
 \begin{equation}
  \frac{\sigma(\gamma\gamma\to K^+K^-)}{\sigma(\gamma\gamma\to K_SK_S)}
  \equiv 2\frac{\sigma(\gamma\gamma\to K^+K^-)}{\sigma(\gamma\gamma\to
  K^0\overline{K^0})}=2\left(\frac{4+\delta}{1+\delta}\right)^2.
  \label{eqn_KpKm_KsKs_ratio}
 \end{equation}
  The conventional charge-counting argument \cite{handbag} corresponds
to $\delta=1$ and therefore gives $25/2$. On the other hand, the
complete suppression of $f'$ gives rise to $32$. Since physically we
expect $\delta$ to be positive definite, we obtain the following
inequality:
 \begin{equation}
  12.5<
  \frac{\sigma(\gamma\gamma\to K^+K^-)}{\sigma(\gamma\gamma\to K_SK_S)}
  < 32.
 \end{equation}
  The upper limit of this equation seems to be satisfied by the 
currently available data \cite{belleks}, within the statistical errors.  
The closeness of the observed ratio at high energy with the limiting 
value $32$ indicates the suppression of the strangeness coupling in this 
region.
  The lower limit seems to be violated at lower energies \cite{belleks}.  
One possible reason is that, in this region, there is dominant 
contribution from $f'$ corresponding to the possibility $\delta>1$, but 
another possible reason is that, as we shall argue later, long-distance 
interaction tends to respect isospin invariance, and so the $a$ coupling 
is suppressed.

  Another quantity that may be interesting is the ratio of the $K^+K^-$ 
and $\pi^+\pi^-$ cross sections. We obtain:
 \begin{equation}
  \frac{\sigma(\gamma\gamma\to K^+K^-)}
  {\sigma(\gamma\gamma\to \pi^+\pi^-)}
  =\left(\frac{4+\delta}{5}\right)^2.
 \end{equation}
  We therefore expect that sufficiently above the threshold, the
$K^+K^-$ cross section is smaller than the $\pi^+\pi^-$ cross section by
up to a factor of $16/25$.

  If the suppression factor $\delta$ is indicative only of the resonance 
structure denoted by $F(f')$, and not the coupling, of $f'$, it then 
follows that
  the $\delta$ obtained by the comparison of these cross sections should 
be approximately the same as the $\delta$ obtained by the comparison of 
the $K^+K^-$ and $K_SK_S$ in the same kinematic range, i.e., the same 
$W_{\gamma\gamma}$ and $\cos\theta^*$ intervals. This should be tested 
experimentally.

 \subsection{Baryon pair production}

  Baryons belong to the $\mathbf3\otimes\mathbf3\otimes\mathbf3$
representation, which is decomposed as:
 \begin{equation}
  \mathbf3\otimes\mathbf3\otimes\mathbf3=
  \mathbf{10_S}\oplus\mathbf{8_{MS}}\oplus\mathbf{8_{MA}}\oplus
  \mathbf{1_A}.
 \end{equation}
  Out of these, phenomenologically the most relevant are the $1/2^+$
baryons in the octet representation.
  The subscripts $\mathbf{MS}$ and $\mathbf{MA}$ stand for mixed
symmetric and mixed antisymmetric, respectively.
  We write them as:
 \begin{eqnarray}
  \mathbf{8_{MS}}&=&B_{(a,b)c}, \\
  \mathbf{8_{MA}}&=&B_{[a,b]c}.
 \end{eqnarray}
  Curly brackets in the subscript represent the symmetric sum and square
brackets represent the antisymmetric sum, so that:
 \begin{equation}
  B_{(a,b)c}=B_{(b,a)c},\qquad B_{[a,b]c}=-B_{[b,a]c}.
 \end{equation}
  Furthermore, $B_{[a,b]c}$ also satisfy the Jacobi identity:
 \begin{equation}
  B_{[a,b]c}+B_{[b,c]a}+B_{[c,a]b}=0.
 \end{equation}
  The symmetric and antisymmetric representations contain the same
physical states, and they are related by:
 \begin{equation}
  B^a_b=\epsilon^{acd}B_{(d,b)c}=\frac12\epsilon^{acd}B_{[c,d]b},
 \end{equation}
  where $\epsilon^{acd}$ is the Levi-Civita tensor with the convention
$\epsilon^{123}=1$. $B^a_b$ is the octet matrix. We first write down the
content of $B_{[a,b]c}$ explicitly:
 \begin{eqnarray}
  &&B_{[1,2]1} = p, \quad
  B_{[1,2]2} = n, \quad B_{[1,2]3} = -2\Lambda/\sqrt6, 
  \nonumber\\
  &&B_{[1,3]1} = -\Sigma^+, 
  B_{[1,3]2} = \Sigma^0/\sqrt2-\Lambda/\sqrt6, B_{[1,3]3} = -\Xi^0,
  \nonumber\\
  &&B_{[2,3]1} = \Sigma^0/\sqrt2+\Lambda/\sqrt6,
  B_{[2,3]2} = \Sigma^-, B_{[2,3]3} = -\Xi^-.
 \end{eqnarray}
  The octet matrix $B^a_b$ is then:
 \begin{equation}
  B^a_b=\left(\begin{array}{ccc}
   \Sigma^0/\sqrt2+\Lambda/\sqrt6 & \Sigma^+ & p \\
   \Sigma^- & -\Sigma^0/\sqrt2+\Lambda/\sqrt6 & n \\
   -\Xi^- & \Xi^0 & -2\Lambda/\sqrt6
  \end{array}\right).
 \end{equation}

  There are three possibilities for evaluating the baryon-baryon-meson 
$SBB$ coupling. We can work in terms of $B_{(a,b)c}$, $B_{[a,b]c}$ or 
$B^a_b$. Here we adopt the following notation:
 \begin{equation}
  \frac12\alpha\overline{B}^{[c,d]a}B_{[c,d]b}M^b_a+
  \beta\overline{B}^{[a,c]d}B_{[b,c]d}M^b_a.
 \end{equation}
  $\alpha$ and $\beta$ are adjustable parameters satisfying
\cite{Nagels:1979xh}:
 \begin{equation}
  \alpha\approx5\beta.
  \label{eqn_alpha_beta_relation}
 \end{equation}
  This relation comes from the approximate flavour $SU(3)$ symmetry for
baryon-meson strong coupling constants.
  On the other hand, for the naive charge-counting argument to work, we
need to impose $\alpha=\beta$. This point will be demonstrated by an
example later.

  We can now tabulate the relevant coupling constants in terms of
$\alpha$ and $\beta$, and these are listed in tab.~\ref{tab_SBB}.
 \begin{table*}[ht]{
  \begin{center}\begin{tabular}{cccc}
   \hline
   final state & $a_0$ & $f_0$ & $f_0'$ \\
   \hline
    $p\bar p$ & $\alpha/\sqrt2$ & $(\alpha+2\beta)\cos\theta_S/\sqrt2$
     & $-(\alpha+2\beta)\sin\theta_S/\sqrt2$ \\
    $n\bar n$ & $-\alpha/\sqrt2$ & $(\alpha+2\beta)\cos\theta_S/\sqrt2$
     & $-(\alpha+2\beta)\sin\theta_S/\sqrt2$ \\
    $\Sigma^+\overline{\Sigma^+}$ & $\frac{\alpha+\beta}{\sqrt2}$
     & $(\alpha+\beta)\cos\theta_S/\sqrt2+\beta\sin\theta_S$
     & $-(\alpha+\beta)\sin\theta_S/\sqrt2+\beta\cos\theta_S$ \\
    $\Sigma^0\overline{\Sigma^0}$ & $0$
     & $(\alpha+\beta)\cos\theta_S/\sqrt2+\beta\sin\theta_S$
     & $-(\alpha+\beta)\sin\theta_S/\sqrt2+\beta\cos\theta_S$ \\
    $\Sigma^-\overline{\Sigma^-}$ & $-\frac{\alpha+\beta}{\sqrt2}$
     & $(\alpha+\beta)\cos\theta_S/\sqrt2+\beta\sin\theta_S$
     & $-(\alpha+\beta)\sin\theta_S/\sqrt2+\beta\cos\theta_S$ \\
    $\Lambda\overline{\Lambda}$ & $0$
     &$\frac{\alpha+5\beta}{3\sqrt2}\cos\theta_S
     +\frac{2\alpha+\beta}3\sin\theta_S$
     &$-\frac{\alpha+5\beta}{3\sqrt2}\sin\theta_S
     +\frac{2\alpha+\beta}3\cos\theta_S$ \\
    $\Xi^0\overline{\Xi^0}$ & $\beta/\sqrt2$
     &$(\alpha+\beta)\sin\theta_S+\beta\cos\theta_S/\sqrt2$
     &$(\alpha+\beta)\cos\theta_S-\beta\sin\theta_S/\sqrt2$ \\
    $\Xi^-\overline{\Xi^-}$ & $-\beta/\sqrt2$
     &$(\alpha+\beta)\sin\theta_S+\beta\cos\theta_S/\sqrt2$
     &$(\alpha+\beta)\cos\theta_S-\beta\sin\theta_S/\sqrt2$ \\
    $\Sigma^0\overline{\Lambda},\Lambda\overline{\Sigma^0}$
     & $\frac{\alpha-\beta}{\sqrt6}$ & $0$ & $0$ \\
   \hline
  \end{tabular}\end{center}
  \caption{The $SB\overline{B}$ coupling coefficients.
  \label{tab_SBB}
  }
 }\end{table*}

  We can consider forming equalities similar to 
eqns.~(\ref{eqn_SPP_sumrule_a}) and (\ref{eqn_SPP_sumrule_b}). In 
particular, we can make use of the ratios of $a_0$ couplings:
 \begin{eqnarray}
  &&\mathcal A(\gamma\gamma\to p\bar p)-
  \mathcal A(\gamma\gamma\to n\bar n):\nonumber\\&&\quad
  \mathcal A(\gamma\gamma\to \Sigma^+\overline{\Sigma^+})-
  \mathcal A(\gamma\gamma\to \Sigma^-\overline{\Sigma^-}):
  \nonumber\\&&\quad
  \mathcal A(\gamma\gamma\to \Xi^0\overline{\Xi^0})-
  \mathcal A(\gamma\gamma\to \Xi^-\overline{\Xi^-}):
  \nonumber\\&&\quad
  \mathcal A(\gamma\gamma\to \Sigma^0\overline{\Lambda})
  \nonumber\\&&\qquad=
  \alpha:\alpha+\beta:\beta:\frac{\alpha-\beta}{2\sqrt{3}}.
 \end{eqnarray}
  However, these will be difficult to verify experimentally. This is 
partly because some final states, for example $n\bar n$, are difficult 
to measure, and partly because we do not expect in any of the pairs of 
reactions above that either of the two amplitudes would become 
sufficiently dominant over the other that the other can be neglected.

  It is hence more helpful to make an estimation analogous to 
eqn.~(\ref{eqn_PP_amp_ratios}).
  For example, the $p\bar p$ amplitude is given, for $\theta_V=0$, by:
 \begin{eqnarray}
  &&\mathcal A(\gamma\gamma\to p\bar p)\nonumber\\&&\qquad=
  \frac5{18}(\alpha+2\beta)\left[\cos^2\theta_SF(f_0)+
  \delta\sin^2\theta_SF(f_0')\right]\nonumber\\&&
  \qquad+\frac\alpha6F(a_0).
 \end{eqnarray}
  As before, by taking $\theta_S=0$, $F(f_0)=F(a_0)$ and absorbing the
difference between $F(f_0)$ and $F(f_0')$ into the coefficient $\delta$,
we arrive at:
 \begin{equation}
  \mathcal A(\gamma\gamma\to p\bar p)\approx
   \frac{F}{9}(4\alpha+5\beta).
 \end{equation}
  We repeat the same exercise for the other production modes and obtain 
results listed in tab.~\ref{tab_gamgam_BB_approx}.
 \begin{table*}[ht]{
  \begin{center}\begin{tabular}{cccc}
   \hline
   final state & $18\mathcal A/F$ & $\alpha=5, \beta=1, \delta=1$ &
   $\delta=0$ \\   \hline
    $p\bar p$ & $8\alpha+10\beta$ & $50$ & $50$ \\
    $n\bar n$ & $2\alpha+10\beta$ & $20$ & $20$ \\
    $\Sigma^+\overline{\Sigma^+}$ &
     $8\alpha+(8+2\delta)\beta$ & $50$ & $48$ \\
    $\Sigma^0\overline{\Sigma^0}$ &
     $5\alpha+(5+2\delta)\beta$ & $32$ & $30$ \\
    $\Sigma^-\overline{\Sigma^-}$ &
     $2\alpha+(2+2\delta)\beta$ & $14$ & $12$ \\
    $\Lambda\overline{\Lambda}$ &
     $(5+4\delta)\alpha/3+(25+2\delta)\beta/3$ & $24$ & $16.67$ \\
    $\Xi^0\overline{\Xi^0}$ &
     $2\delta\alpha+(8+2\delta)\beta$ & $20$ & $8$ \\
    $\Xi^-\overline{\Xi^-}$ &
     $2\delta\alpha+(2+2\delta)\beta$ & $14$ & $2$ \\
    $\Sigma^0\overline{\Lambda},\Lambda\overline{\Sigma^0}$ &
     $(\alpha-\beta)\sqrt3$ & $6.93$ & $6.93$ \\
   \hline
  \end{tabular}\end{center}
  \caption{The limiting behaviour of $\gamma\gamma\to B\overline B$
amplitudes.
  \label{tab_gamgam_BB_approx}
  }
 }\end{table*}

  The statement made earlier about naive charge-counting in the case 
$\alpha=\beta$ can now be demonstrated explicitly. For example, the 
ratio of the $p\bar p$ and $n\bar n$ amplitudes is given by:
 \begin{equation}
  \frac{8+10}{2+10}=\frac32=\frac{1^2+2^2+2^2}{1^2+1^2+2^2}.
 \end{equation}

  The ratios of cross sections sufficiently above the threshold region 
are given by the square of the coefficients listed in 
tab.~\ref{tab_gamgam_BB_approx}, so that, for instance:
 \begin{equation}
  \frac{\sigma(\gamma\gamma\to p\bar p)}{
        \sigma(\gamma\gamma\to\Sigma^0\overline{\Sigma^0})}
  \approx\left(\frac{8\alpha+10\beta}{5\alpha+(5+2\delta)\beta}\right)^2
  .
 \end{equation}
  This quantity comes out to be between $\sim2.4$ and $\sim2.7$ for 
$\alpha=5\beta$ and $0<\delta<1$. The measurement of the 
$\Xi^0\overline{\Xi^0}$, or $\Xi^-\overline{\Xi^-}$, cross section would 
be particularly interesting because of the sensitivity to $\delta$. From 
the discussion of pseudo-scalar meson pair production, we expect that 
$\delta$ is small, so that this cross section would be suppressed by 
factor $\sim(50/8)^2\sim40$ compared to the $p\bar p$ cross section.

 \section{Long- and short-distance dynamics} \label{sec_dynamics}

  Let us model the short-distance amplitude as a scaling contribution:
 \begin{equation}
  \mathcal A\propto \frac{s^{4-K/2}}{(t-M^2_1)(u-M^2_2)}.
  \label{eqn_scaling_angular_distribution}
 \end{equation}
  $K$ is as appearing in eqn.~(\ref{eqn_quark_counting}).
  We take the two masses $M_1$ and $M_2$ to be the corresponding hadron 
masses as opposed to, for instance, some appropriate quark masses. The 
contribution of these mass terms is negligible in any case in the region 
of interest. This expression gives rise to the angular distribution 
$\approx(1-\cos^2\theta^2)^{-2}$ which is characteristic of single quark 
exchange, or more generally light particle exchange, in the $t$-channel.  
This angular distribution would be valid at high-energy, and the scaling 
behaviour of eqn.~(\ref{eqn_quark_counting}) implies that the angular 
distribution must remain the same at low-energy.

  To this, we add a long-distance pole(-resonance) contribution which 
has the Regge limiting form:
 \begin{eqnarray}
  \mathcal A&\propto&\Gamma(\ell-\alpha(t))
   \left(1+\tau\exp(-i\pi\alpha(t))\right)
   (\alpha's)^{\alpha(t)} \nonumber\\
  &+&\left(t\leftrightarrow u\right),
  \label{eqn_regge_general}
 \end{eqnarray}
  The linear trajectory is parametrized $\alpha(t)=\alpha(0)+\alpha't$ 
as usual, with $\alpha'\approx0.9$~GeV$^{-2}$.
  $\ell$ is the lowest spin of the trajectory, and $\tau=\pm1$ is the 
signature. For baryons, the signature term is modified to 
\cite{storrow}:
 \begin{equation}
  1+\tau\exp\left(-i\pi\left(\alpha(t)-1/2\right)\right).
  \label{eqn_regge_signature_baryon}
 \end{equation}

  It is found \cite{odagiriveneziano} that a simple Regge expression 
similar to eqn.~(\ref{eqn_regge_general}) yields characteristic 
behaviour in the central region $\cos\theta^*\approx0$ that is in 
accordance with the $\gamma\gamma\to p\bar p$ data just above the 
threshold. This may seem surprising at first sight, but is reasonable 
considering that the (Regge) pole amplitude and the resonance amplitude 
are related by (semi-local) duality \cite{finite_energy_sumrule}. After 
integrating over the resonances, the behaviour of the two amplitudes is 
similar. In particular, this method works well in the case of $p\bar p$ 
since the cross section is a smooth function of $W_{\gamma\gamma}$ and 
no trace of resonances is seen. For $K^+K^-$, the resonance structure is 
still seen, so that we may, for instance, replace the Regge amplitude 
with a resonance--pole dual amplitude of the Veneziano model 
\cite{veneziano}.

 \subsection{Baryon pair production}

  Let us first consider $p\bar p$ production.
  In eqn.~(\ref{eqn_scaling_angular_distribution}), we set $K=8$. In 
eqn.~(\ref{eqn_regge_general}) with the modification of 
eqn.~(\ref{eqn_regge_signature_baryon}), we set $\ell=\frac12$.
  As for the trajectories, ref.~\cite{storrow}, the leading $S=0$ 
contributions, the $N/\Delta$, have the following parametrization:
 \begin{eqnarray}
  N_\alpha: && \alpha(t)=-0.34+0.99t,\\
  N_\gamma: && \alpha(t)=-0.63+0.89t,\\
  N_\beta: && \alpha(t)=0+0.9t,\\
  \Delta_\delta: && \alpha(t)=0.07+0.92t.
 \end{eqnarray}
  $N_\alpha$ and $N_\beta$ are signature even whereas $N_\gamma$ and
$N_\delta$ are signature odd. $N$ are isospin $1/2$ and $\Delta$ are
isospin $3/2$, as usual. Both exchanges are allowed, although $\Delta$
exchange can only take place in $\rho\rho\to p\bar p$ and not in the
other subprocesses.

  In principle, we should include all four contributions. In practice, 
however, we found phenomenologically that the inclusion of just one 
trajectory, the $N_\beta$ trajectory, is sufficient.

  For the explicit $\alpha'$ in eqn.~(\ref{eqn_regge_general}), as 
opposed to the $\alpha'$ implicit in the trajectory $\alpha(t)$, we 
adopt $0.9$~GeV$^{-2}$.

\begin{figure}[ht]{
 \begin{center}
  $\sigma(\gamma\gamma\to p\bar p)/$nb, $|\cos\theta^*|<0.6$\\
  \epsfig{file=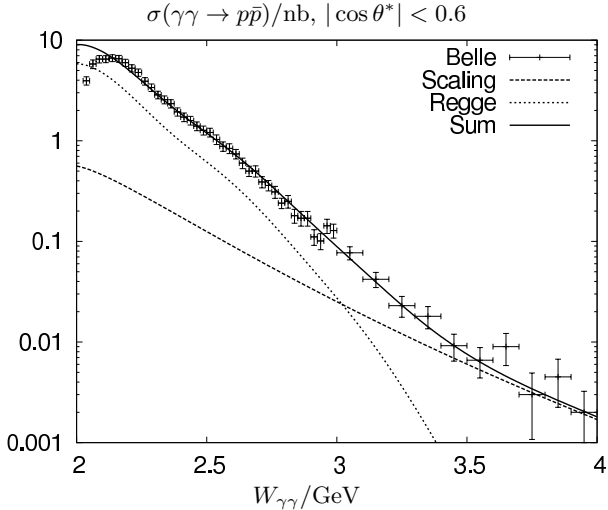,width=8cm}\\
  $W_{\gamma\gamma}/$GeV
 \end{center}
 \caption{The $\gamma\gamma\to p\bar p$ integrated cross section in the 
region $|\cos\theta^*|<0.6$. We show the Belle result against three 
theoretical results: scaling based on the quark-counting rule, the Regge 
amplitude, and the sum of the two.
  The vertical error-bars only includes the statistical uncertainty.
 \label{fig_ppbar_cs}}}\end{figure}

  In fig.~\ref{fig_ppbar_cs}, we show the scaling amplitude, the Regge 
amplitude, and the sum of the two.
  We first fix the normalization of the scaling contribution by fitting 
by the eye with the data at $W_{\gamma\gamma}$ near $4$~GeV. We then 
adjust the Regge contribution so that the sum of the two terms fits the 
integrated cross section.

  There is good agreement with the data, except in the region just above 
the threshold. Even this region shows improvement compared with our 
previous calculation in ref.~\cite{odagiriveneziano}, where the 
signature term was neglected.
  We note that by further modifying the signature term by the artificial 
substitution $\tau\to-i$, we were able to obtain the fall-off near 
threshold seen in the real data. This suggests the possibility that the 
inclusion of other trajectories and/or resonances with appropriate 
strengths may change the threshold behaviour.

\begin{figure*}[ht]{
 \begin{center}
 \epsfig{file=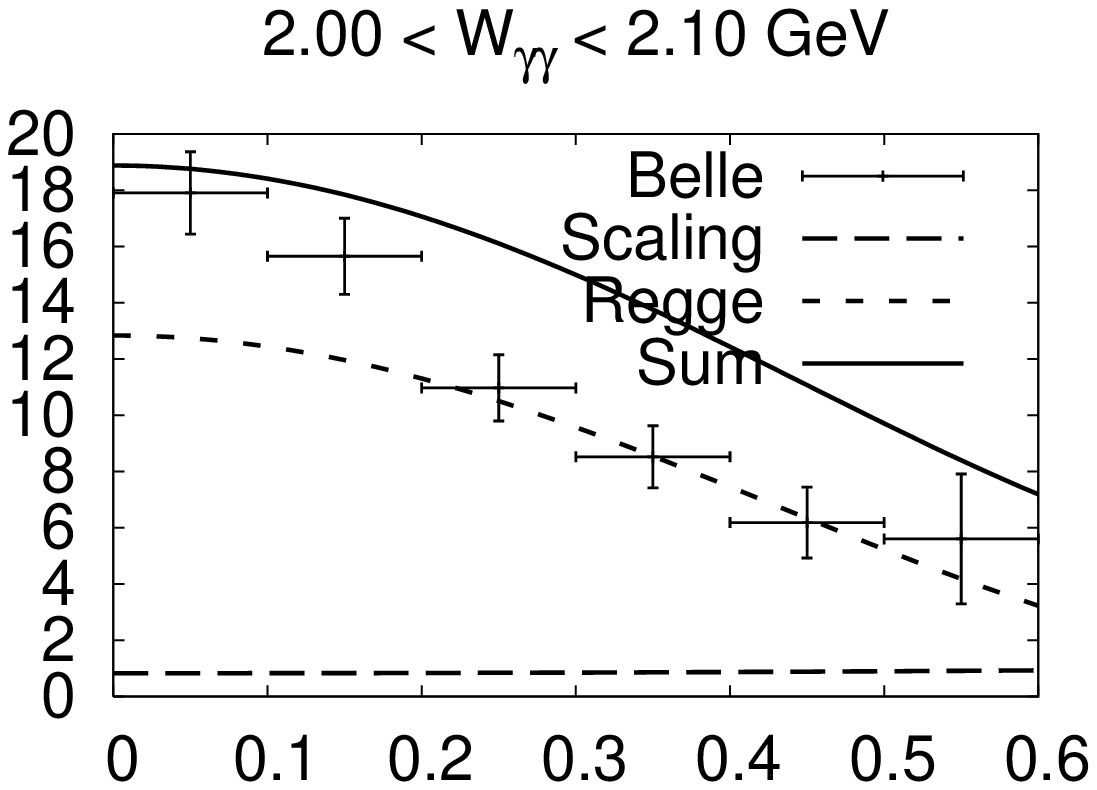,width=4.9cm}
 \epsfig{file=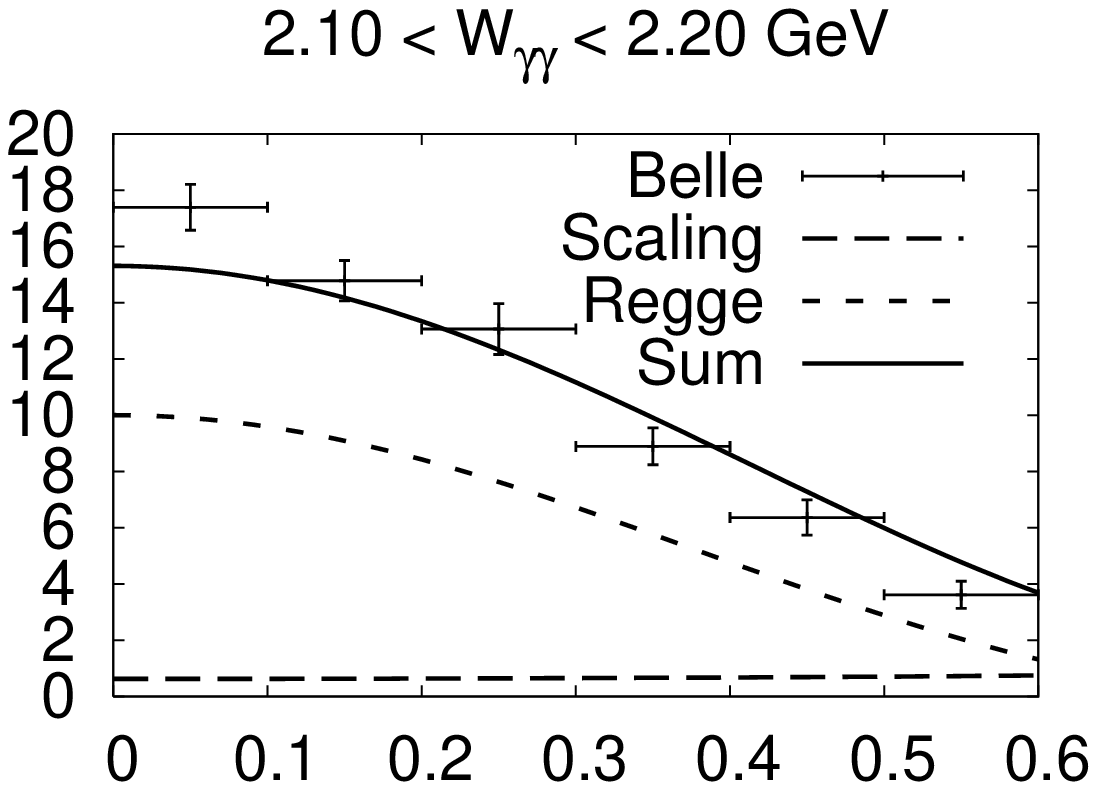,width=4.9cm}
 \epsfig{file=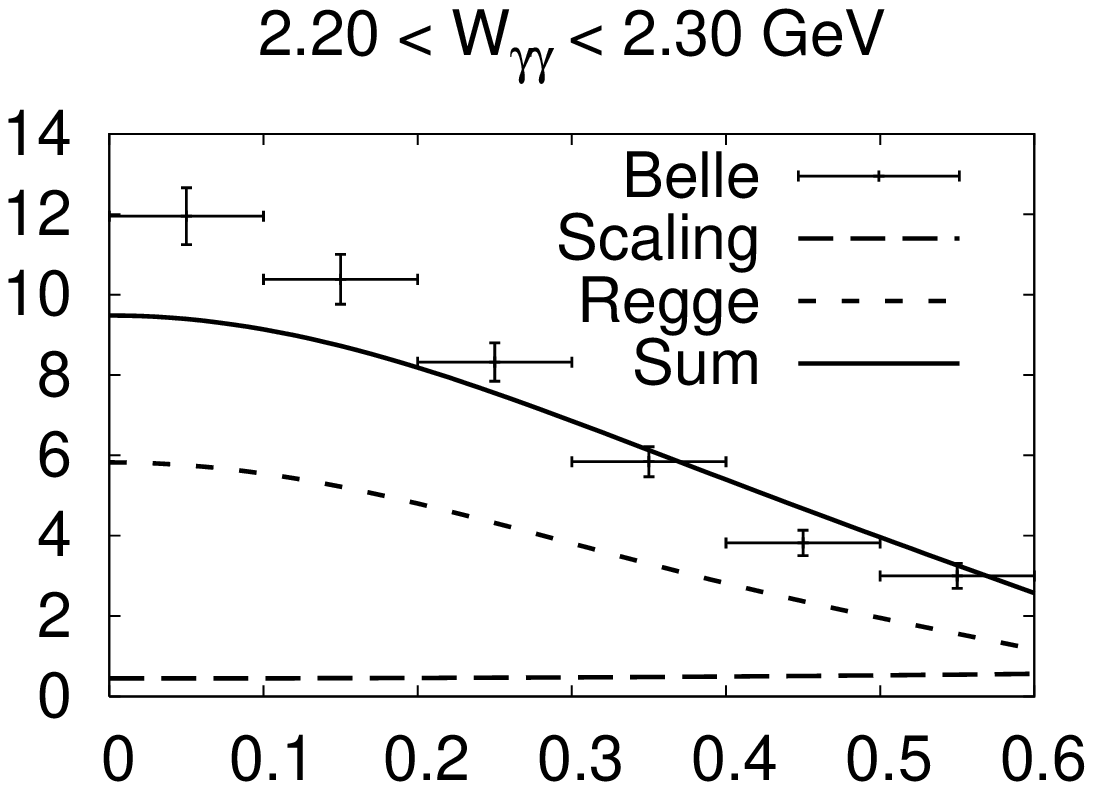,width=4.9cm}

 \epsfig{file=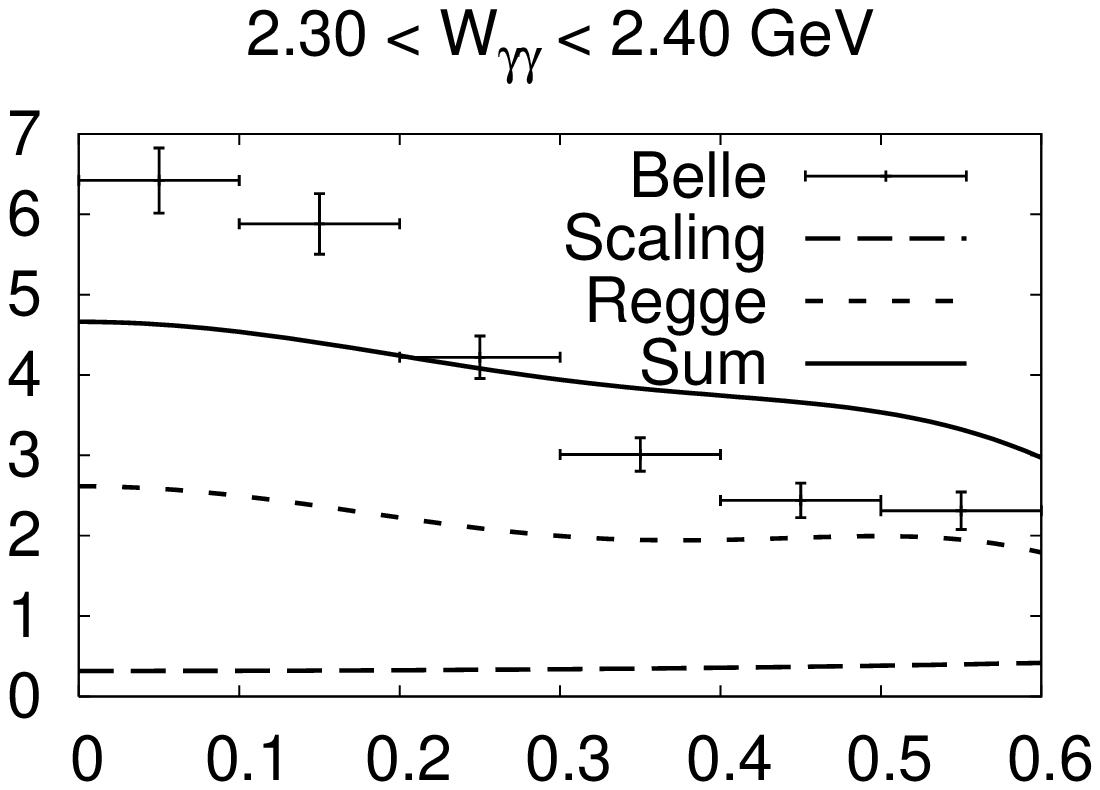,width=4.9cm}
 \epsfig{file=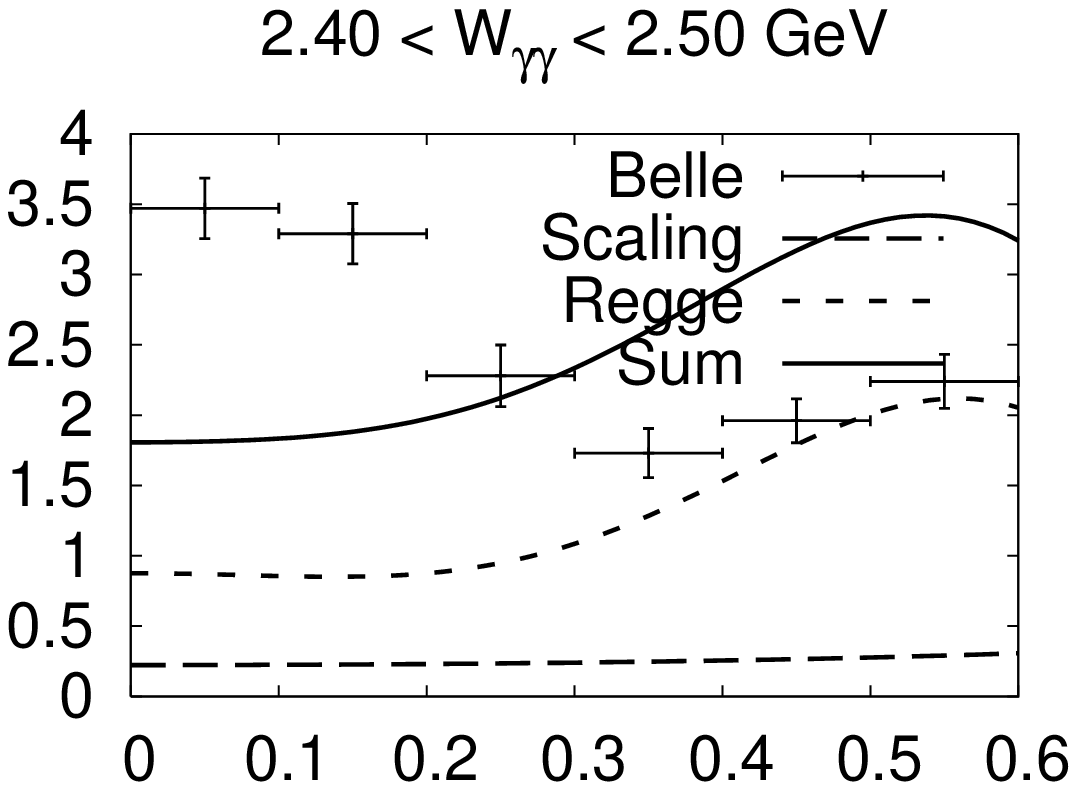,width=4.9cm}
 \epsfig{file=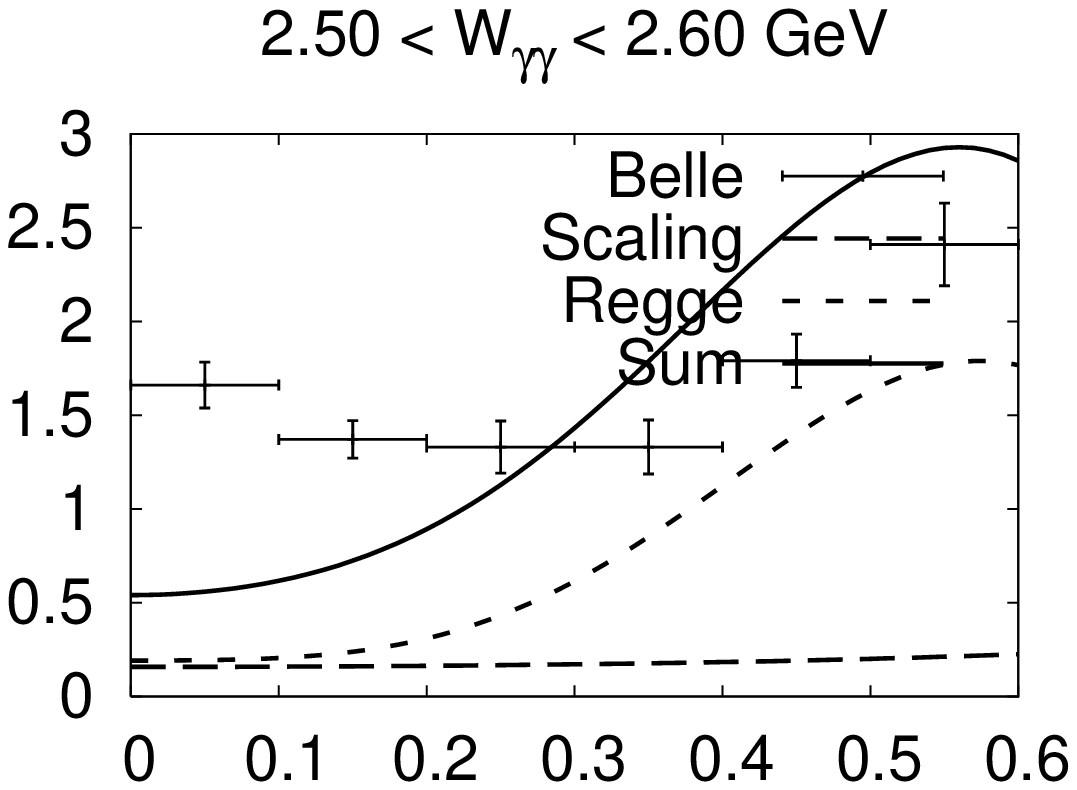,width=4.9cm}

 \epsfig{file=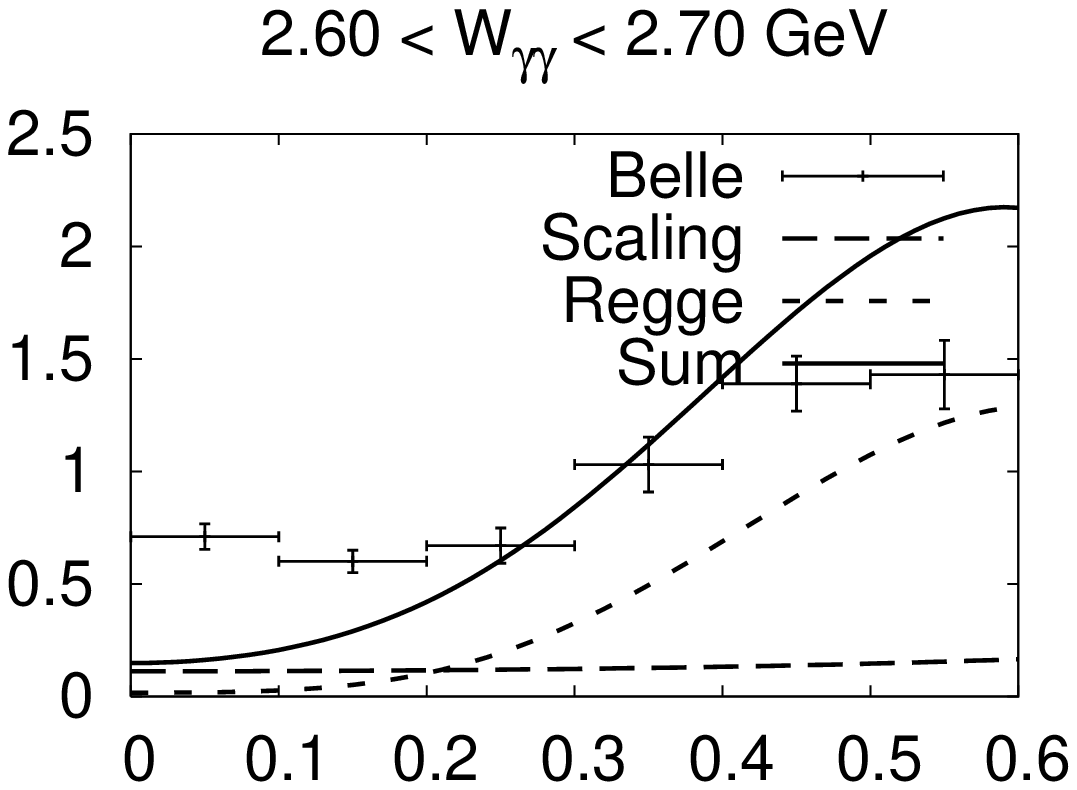,width=4.9cm}
 \epsfig{file=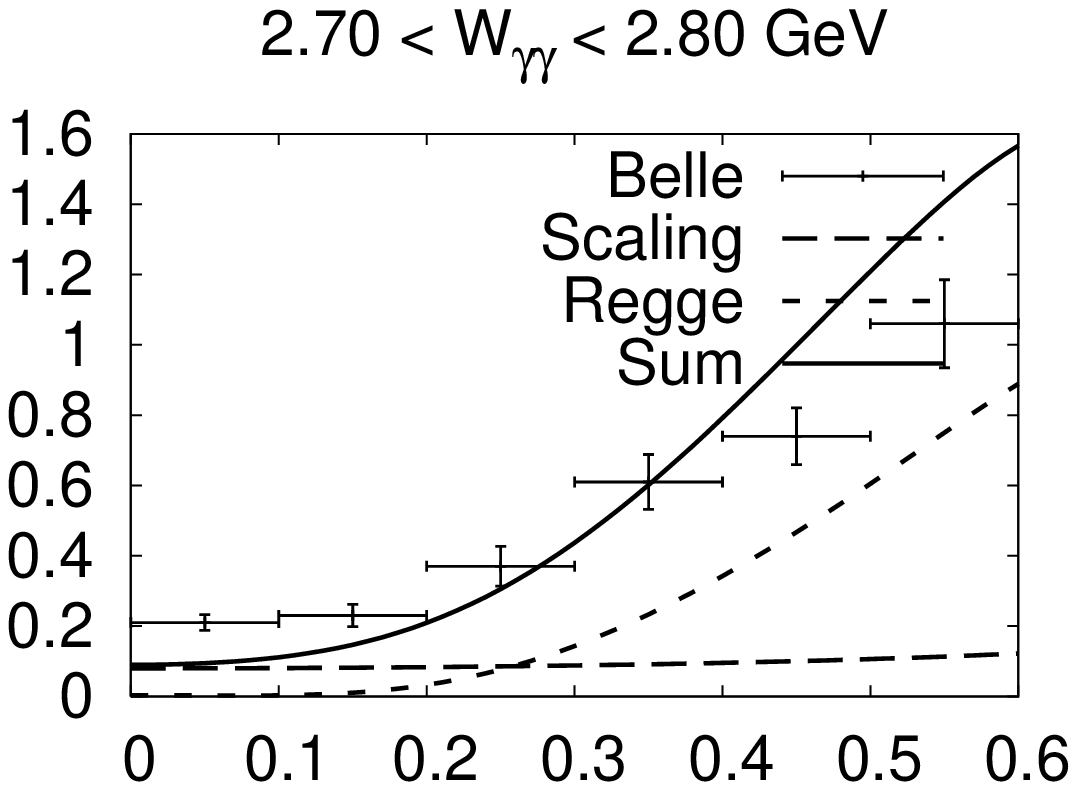,width=4.9cm}
 \epsfig{file=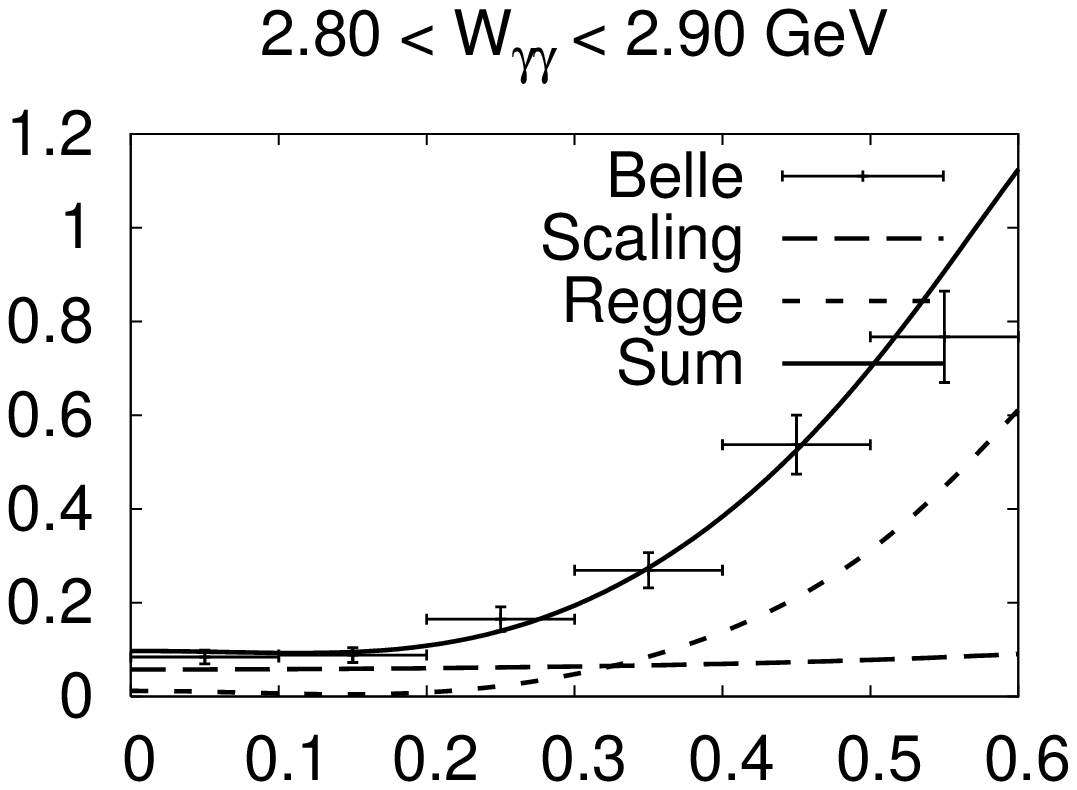,width=4.9cm}

 \epsfig{file=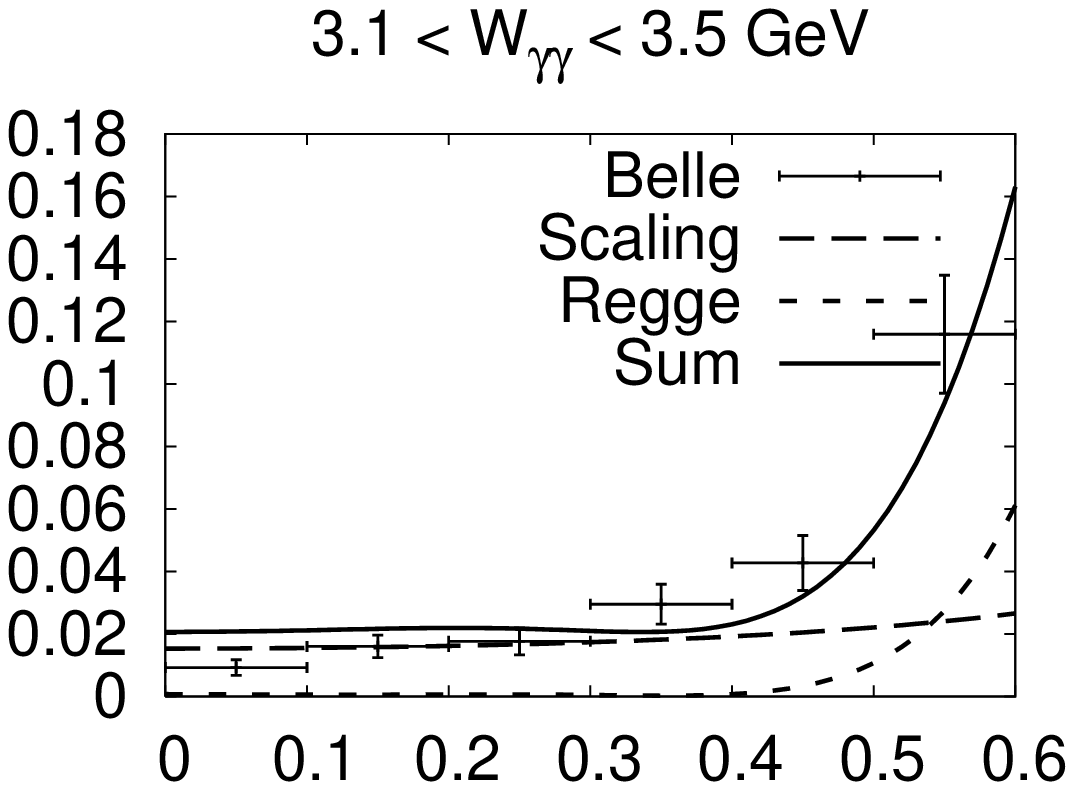,width=4.9cm}
 \epsfig{file=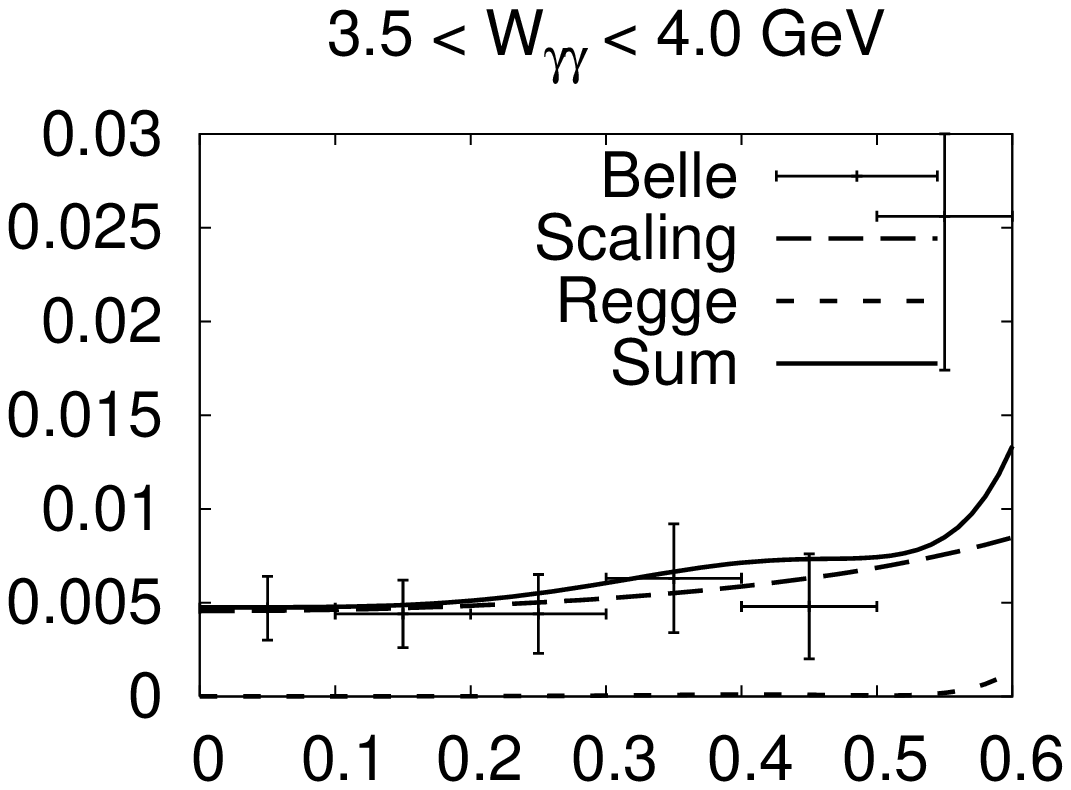,width=4.9cm}
 \end{center}
 \caption{The angular distribution of $\gamma\gamma\to p\bar p$. The
results of three theoretical model calculations are compared against the
experimental data from ref.~\cite{bellep}.
 \label{fig_ppbar_ad}}}\end{figure*}

  The angular distributions are shown in fig.~\ref{fig_ppbar_ad}.
  The shifting of the peak of the central angular distribution from the 
$\cos\theta\approx0$ region to the forward region occurs slightly faster 
(about 100~MeV faster) in the theoretical curve than in the experimental 
data.
  However, the overall trend is in fair agreement with that seen in the 
experiment.

  Having achieved this level of agreement, it becomes desirable to be
able to extend our results to the case of other baryons, for instance 
$\Lambda$ and $\Sigma^0$ \cite{baryon_l3,baryon_cleo}.
  However, there is no good method for estimating the Regge couplings 
\cite{ddln}.
  On the other hand, we expect that for all baryons, the Regge 
contributions dominate over the scaling contribution, since the Regge 
contribution is expected to be more insensitive to the type of the 
baryon \cite{ddln}, whereas the scaling contribution, from 
tab.~\ref{tab_gamgam_BB_approx}, is always smaller than the proton pair 
case.

  As noted in ref.~\cite{odagiriveneziano}, just above the threshold 
region, we expect invariance under $u\leftrightarrow d$, as opposed to 
the $d\leftrightarrow s$ symmetry that follows from the perturbative 
approaches \cite{handbag,berger-schweiger}. This implies, in the 
$s$-channel picture, the suppression of isovector $a$ component. Hence 
$\Sigma^0\overline{\Lambda}$ production would be suppressed.

  The size of the scaling contribution to each baryon pair can be 
estimated from tab.~\ref{tab_gamgam_BB_approx}, but this is, as seen in 
the above results, small.
  The more central, or the higher-energy region, is expected to have a 
more short-distance character and so the argument of 
sec.~\ref{sec_clebschgordan} can be applied.

 \subsection{Meson pair production}

  We now consider meson pair photoproduction.
  In eqn.~(\ref{eqn_scaling_angular_distribution}), we set $K=6$. In 
eqn.~(\ref{eqn_regge_general}), we set $\ell=1$.

  The signature $\tau$ in eqn.~(\ref{eqn_regge_general}) can be $\pm1$ 
depending on the spin of the trajectory. However, unlike in the baryonic 
case discussed above, we have spin-degenerate trajectories with 
$\tau=+1$ and $\tau=-1$. $\tau=+1$ corresponds to the exchange of 
even-spin mesons:
 \begin{equation}
  \mathcal A\propto\Gamma(1-\alpha(t))
   \left(1+\exp(-i\pi\alpha(t))\right)
   (\alpha's)^{\alpha(t)}.
  \label{eqn_regge_even}
 \end{equation}
  The odd-spin mesons have $\tau=-1$ and:
 \begin{equation}
  \mathcal A\propto\Gamma(1-\alpha(t))
   \left(1-\exp(-i\pi\alpha(t))\right)
   (\alpha's)^{\alpha(t)}.
  \label{eqn_regge_odd}
 \end{equation}
  Adding together the contributions of degenerate trajectories, the 
contribution of the signature term in general tends to cancel. In the 
limiting case of perfect cancellation, there are two possibilities:
 \begin{enumerate}
 \item The two amplitude add with the same sign, leading to the 
cancellation of $\exp(-i\pi\alpha(t))$. This corresponds to the 
cancellation of the handbag diagrams in fig.~\ref{fig_qline_meson}.
 \item The two amplitudes add with the opposite sign, leading to the 
cancellation of the constant-phase term. This corresponds to the the 
cancellation of the cat's-ears diagram in fig.~\ref{fig_qline_meson}.
 \end{enumerate}
  Let us denote these respectively as `Regge cat's-ears' and `Regge 
handbag'. It turns out that the plateau structure of the $K^+K^-$ 
integrated cross section is only reproduced in the `Regge handbag' case, 
since a rotating phase is necessary to yield non-trivial interference 
with the scaling contribution.

  Let us therefore consider the `Regge handbag' case.
  Here, we can write the combined amplitude as a $s-t$ dual amplitude. 
Using the simple Veneziano amplitude of ref.~\cite{odagiriveneziano}, we 
are able to simulate both the resonance and the pole regions with the 
expression:
 \begin{equation}
  \mathcal A\propto\frac{\Gamma(1-\alpha(t))\Gamma(1-\alpha(s))}
  {\Gamma(1-\alpha(t)-\alpha(s))} + (u\leftrightarrow t),
 \end{equation}
  from which we can recover the Regge amplitude by the application of 
the Stirling factorial approximation. Because of the resonance--pole 
duality, the discussion of the preceding sec.~\ref{sec_clebschgordan} 
holds. This has the implication that relations between amplitudes such 
as eqns.~(\ref{eqn_SPP_sumrule_a}) and (\ref{eqn_SPP_sumrule_b}) are 
satisfied.

\begin{figure}[ht]{
 \begin{center}
  $\sigma(\gamma\gamma\to K^+K^-)/$nb, $|\cos\theta^*|<0.6$\\
  \epsfig{file=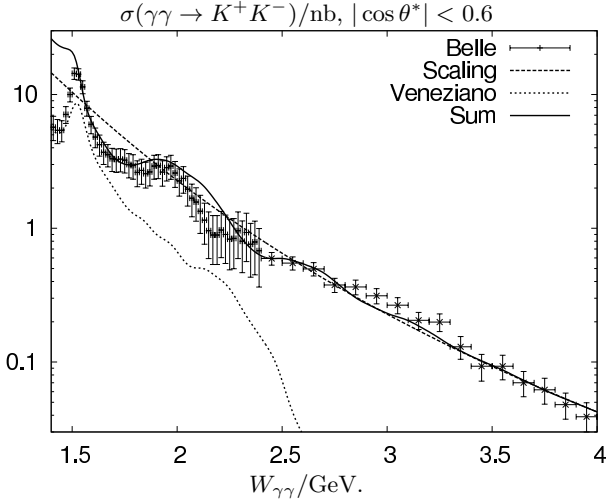,width=8cm}\\
  $W_{\gamma\gamma}/$GeV.
 \end{center}
 \caption{$\gamma\gamma\to K^+K^-$ integrated cross section in the 
region $|\cos\theta^*|<0.6$. The experimental results \cite{bellek} are 
compared against scaling, the Veneziano model, and the sum of the two.
 \label{fig_kpkm_model}}}\end{figure}

  We fit the scaling amplitude by the eye to the data at near $4$~GeV. 
We then study the behaviour of the sum of this amplitude and the 
parametrization of ref.~\cite{odagiriveneziano}. The result is shown in 
fig.~\ref{fig_kpkm_model}.
  The scaling curve fits the data reasonably above about $2.5$~GeV. 
Below $2.5$~GeV, the behaviour of the integrated cross section is still 
close to the scaling curve, but this is accidental since the behaviour 
of the angular distribution is far from that parametrized by 
eqn.~(\ref{eqn_scaling_angular_distribution}).
 On the other hand, the Veneziano amplitude by itself provides a 
semi-quantitative description of the data below about $2$~GeV although 
the plateaus just below and above $2$~GeV are not reproduced.

  The sum of the two amplitudes shows striking resemblance to the real 
data, except below $1.5$~GeV. In particular, this reproduces the 
plateaus. These come from the non-trivial interference between the 
long-distance and short-distance contributions to the amplitude, and 
are, as seen in fig.~\ref{fig_kpkm_model}, not correlated directly with 
the shape of the resonances.

\begin{figure*}[ht]{
 \begin{center}
 \epsfig{file=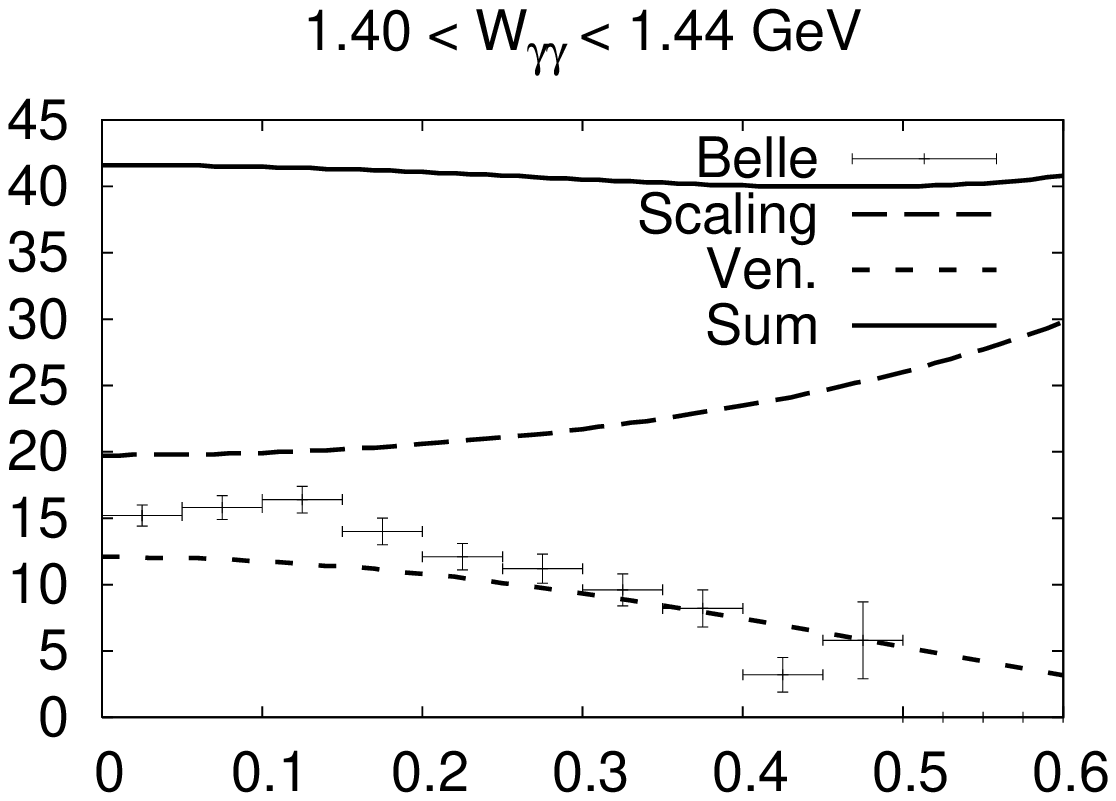,width=5cm}
 \epsfig{file=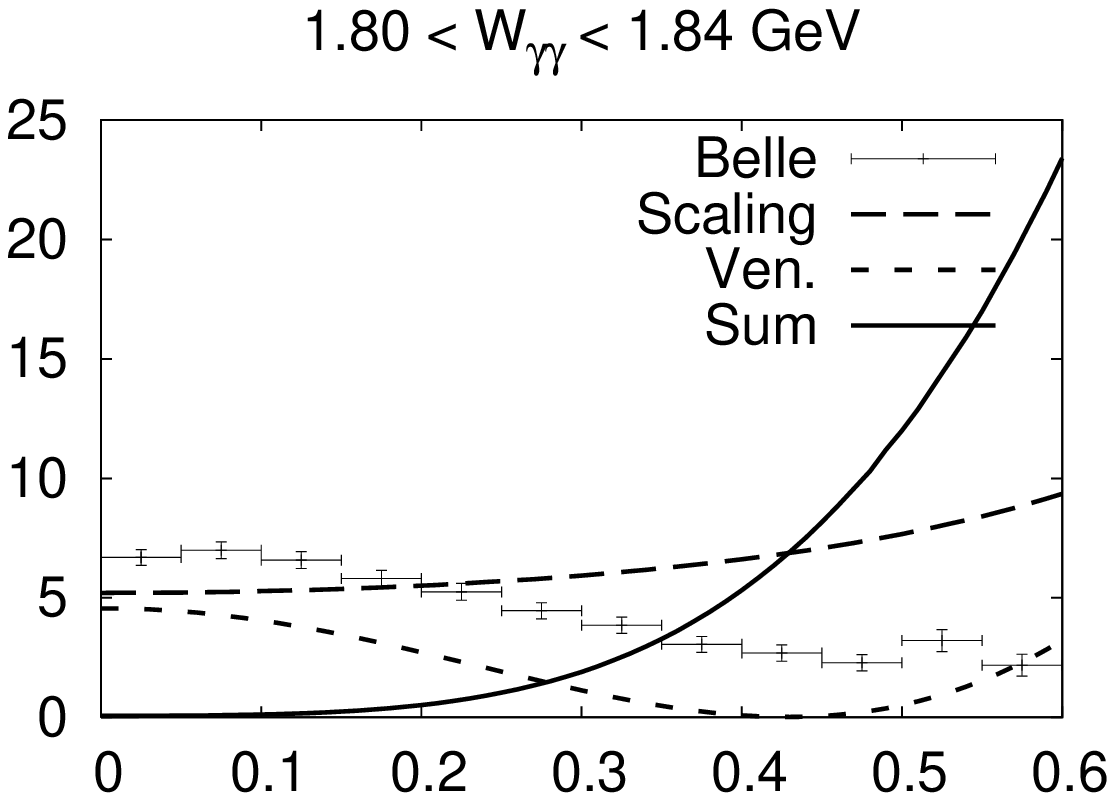,width=5cm}
 \epsfig{file=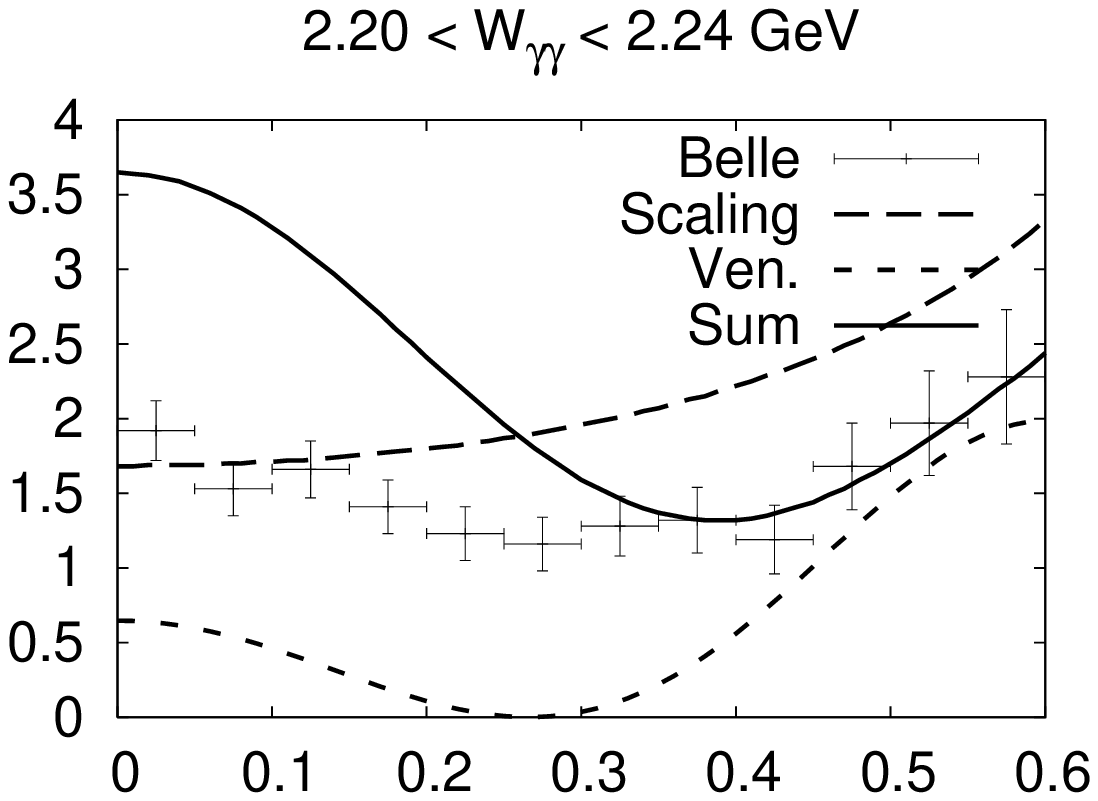,width=5cm}

 \epsfig{file=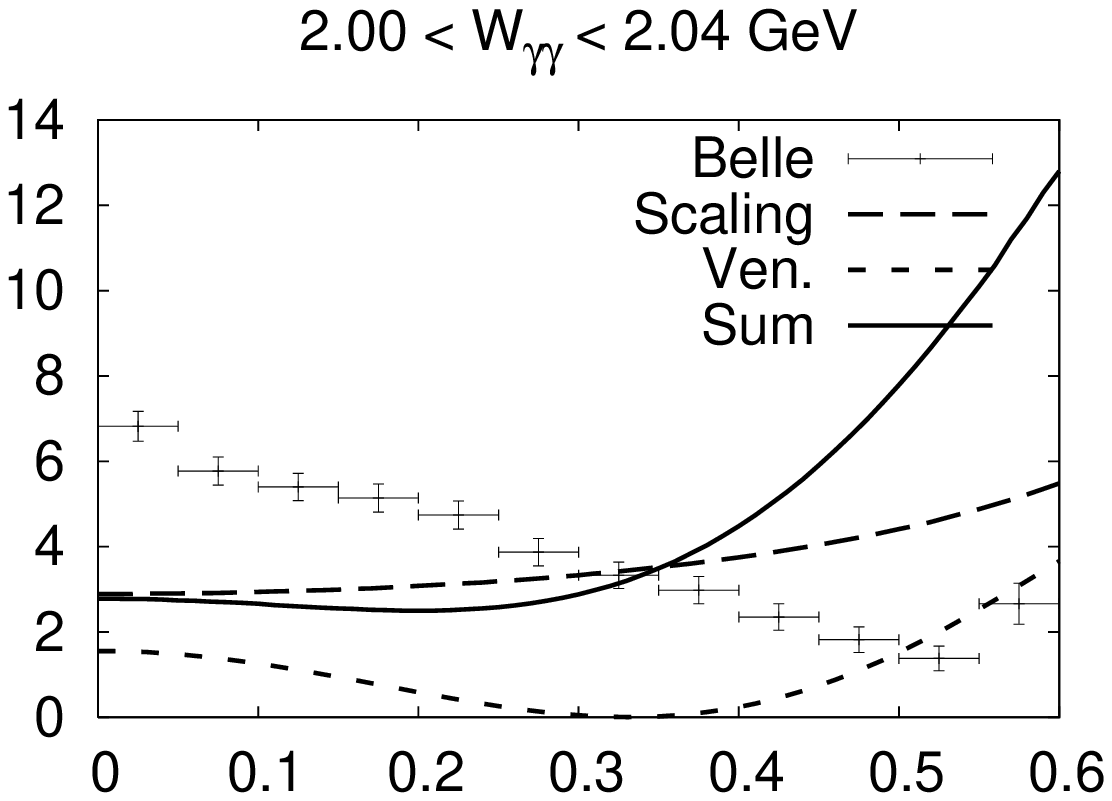,width=5cm}
 \epsfig{file=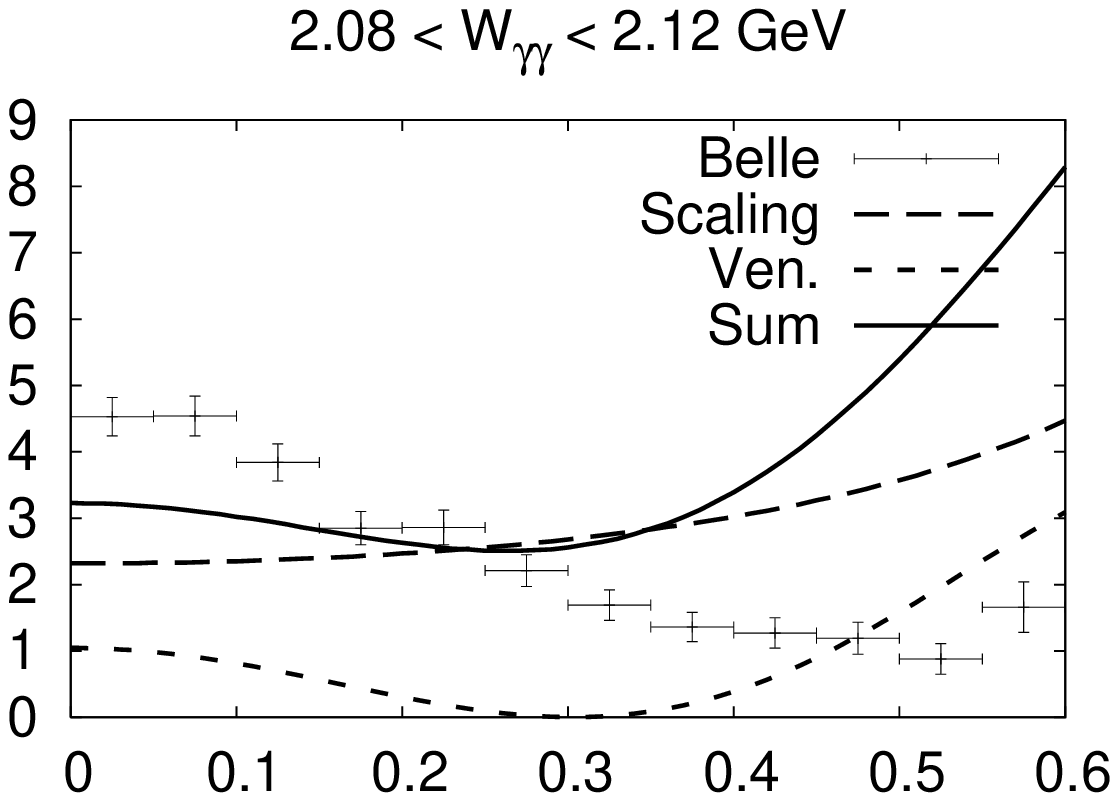,width=5cm}
 \epsfig{file=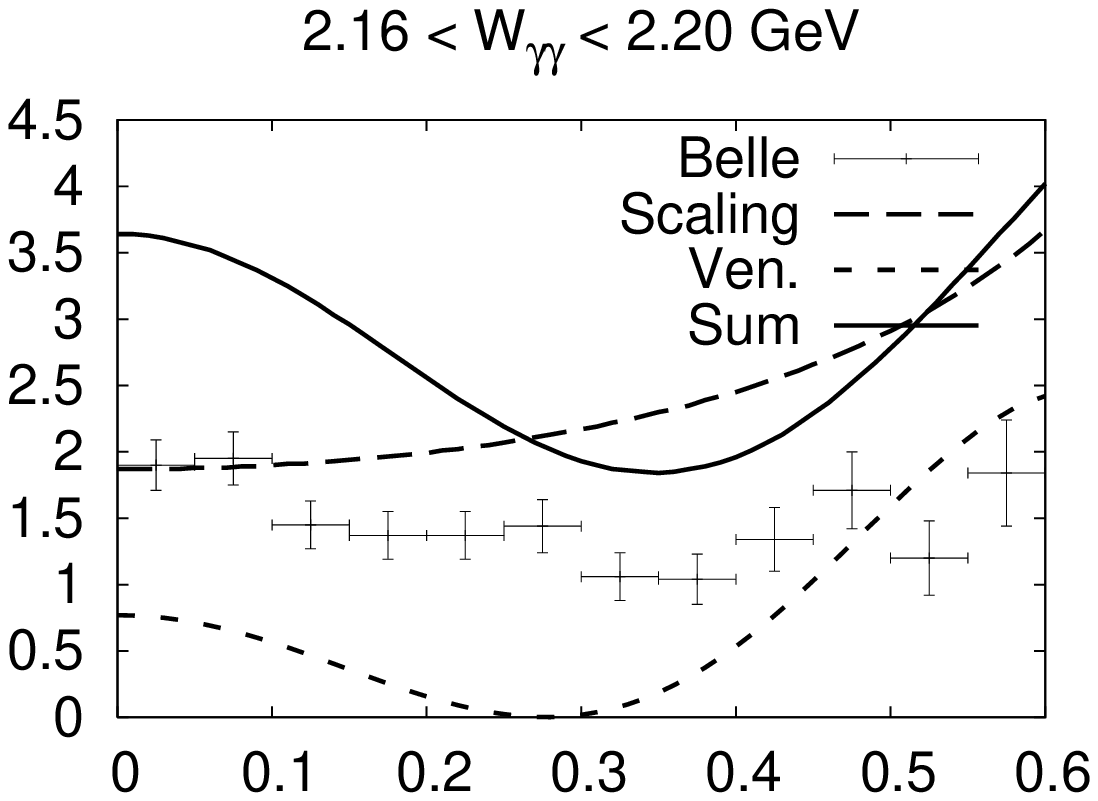,width=5cm}
 \end{center}
 \caption{The angular distribution of $\gamma\gamma\to K^+K^-$ at three 
representative energy ranges in between $1.40$ and $2.24$~GeV (upper 
row), and between $2.00$ and $2.20$~GeV (lower row).
  We show the Belle data, the scaling contribution, the Veneziano-model 
contribution and the squared sum of the two.
  The vertical error-bars on the Belle data are statistical only.
 \label{fig_kpkm_ven_ad}}}\end{figure*}

  This apparent resemblance with the data is, however, misleading, since 
the angular distribution does not correctly reproduce the structure of 
the experimental data. This is seen in fig.~\ref{fig_kpkm_ven_ad}, which 
shows the distribution at three representative energy ranges and in the 
region between $2.00$ and $2.20$ GeV.
  For most of the energy range, the sum of the two amplitudes does not 
yield a better approximation to the angular distribution than either of 
the two individual contributions.

\begin{figure}[ht]{
 \begin{center}
  $\sigma(\gamma\gamma\to K^+K^-)/$nb, $|\cos\theta^*|<0.6$\\
  \epsfig{file=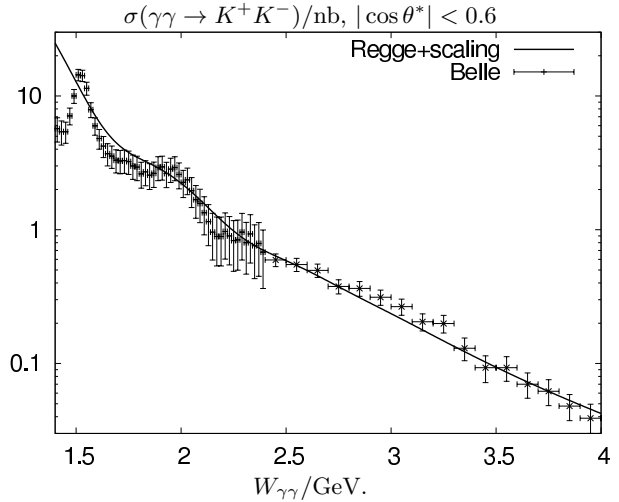,width=8cm}\\
  $W_{\gamma\gamma}/$GeV.
 \end{center}
 \caption{$\gamma\gamma\to K^+K^-$ integrated cross section in the 
region $|\cos\theta^*|<0.6$. The experimental results \cite{bellek} are 
compared against the sum of the Regge and scaling amplitudes.
 \label{fig_kpkm_lowregge}}}\end{figure}

\begin{figure*}[ht]{
 \begin{center}
 \epsfig{file=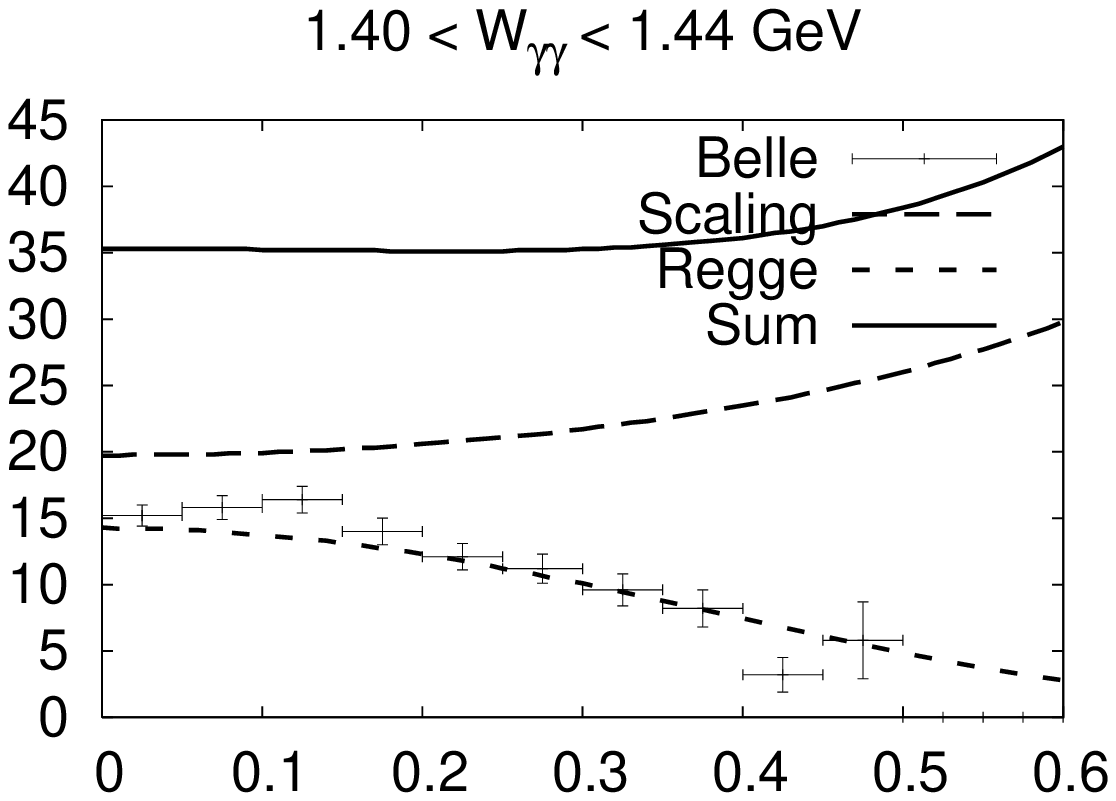,width=5cm}
 \epsfig{file=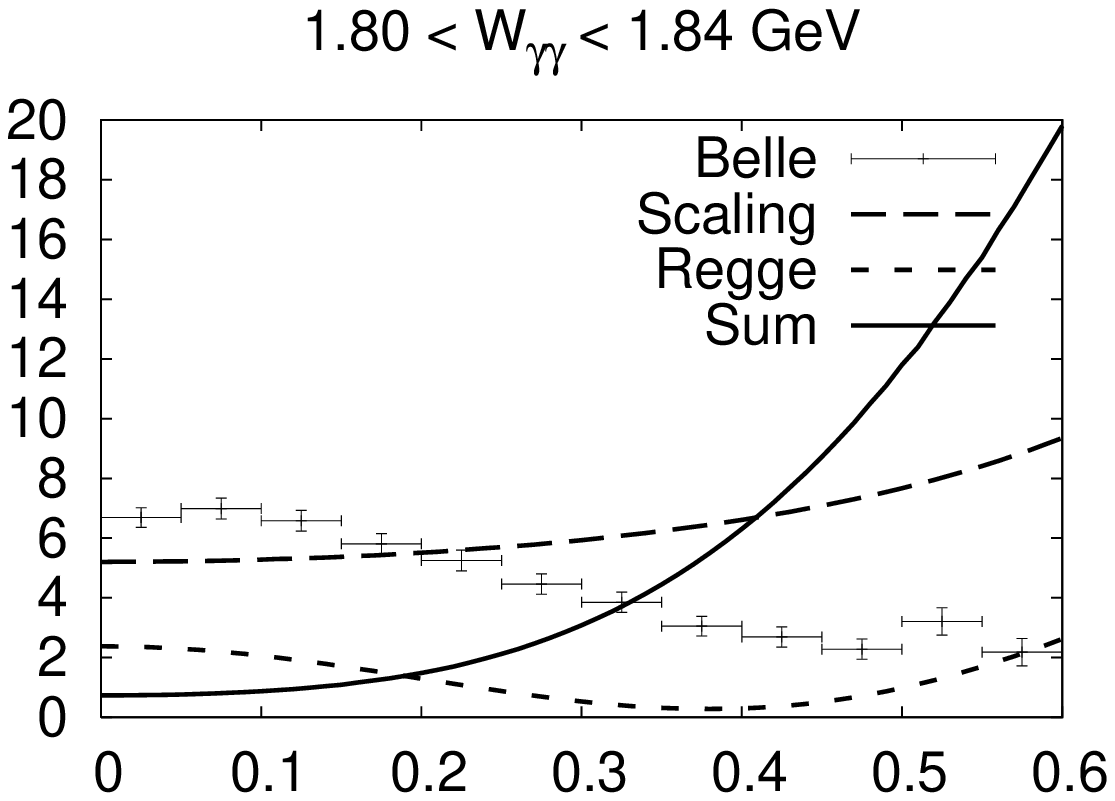,width=5cm}
 \epsfig{file=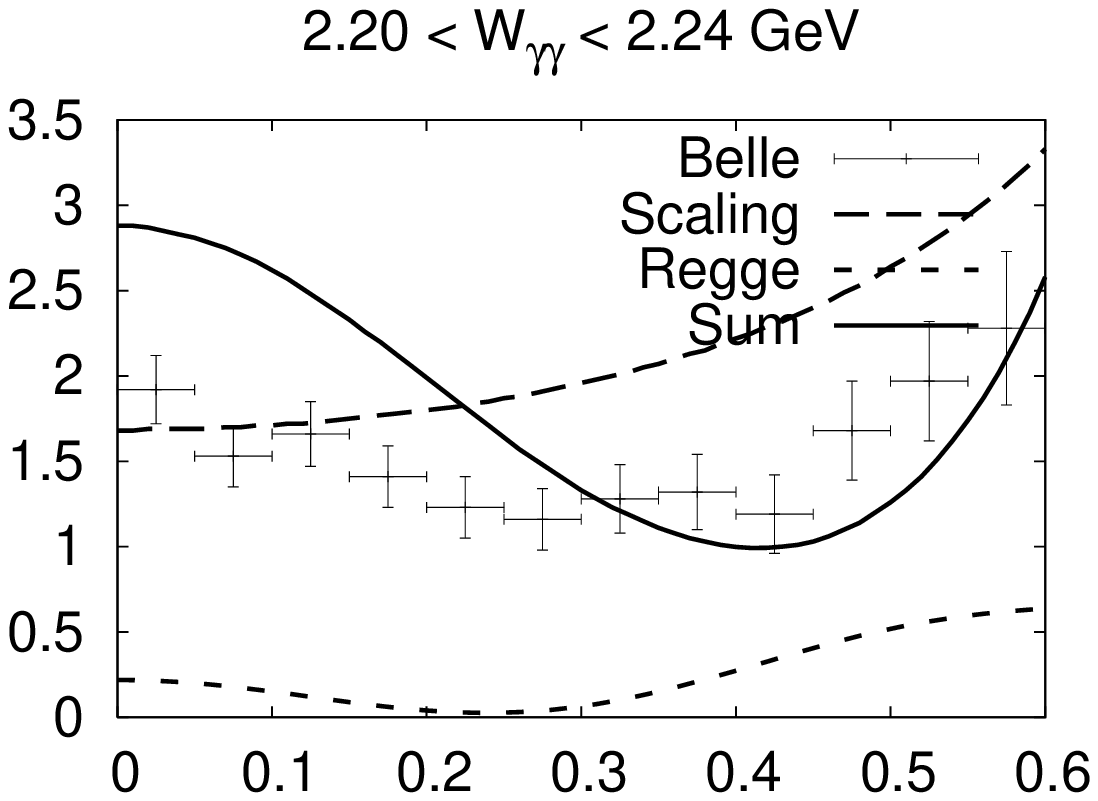,width=5cm}

 \epsfig{file=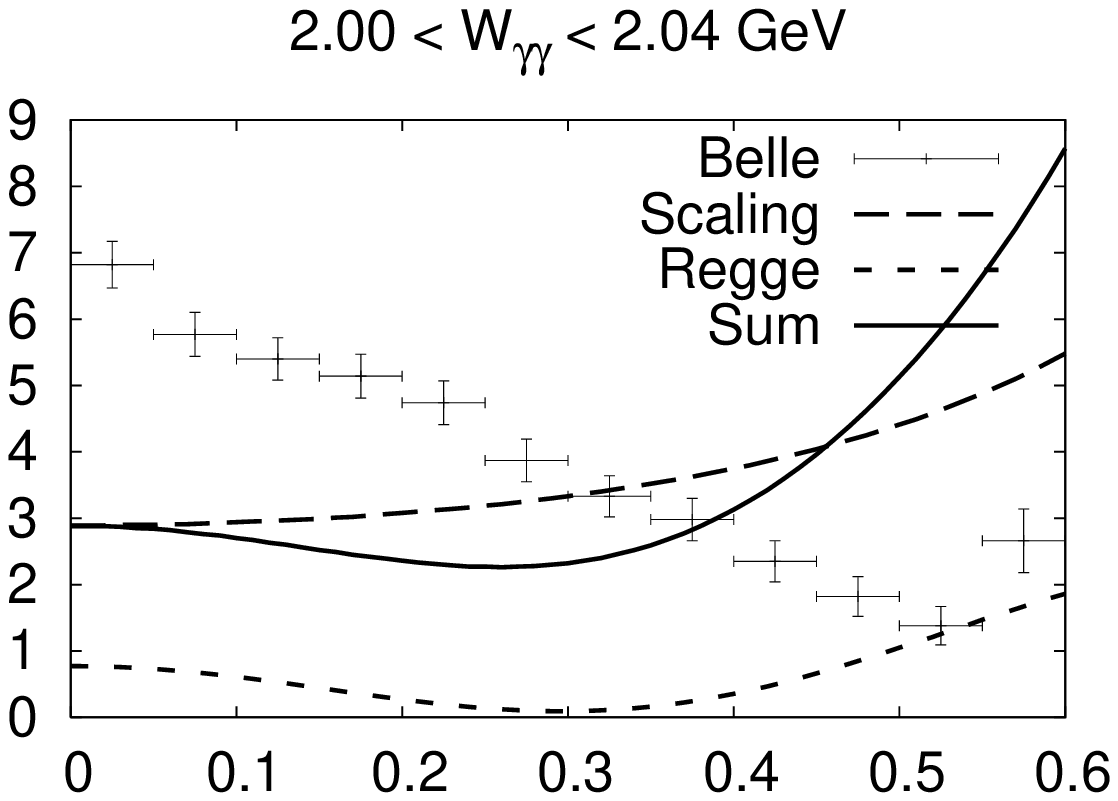,width=5cm}
 \epsfig{file=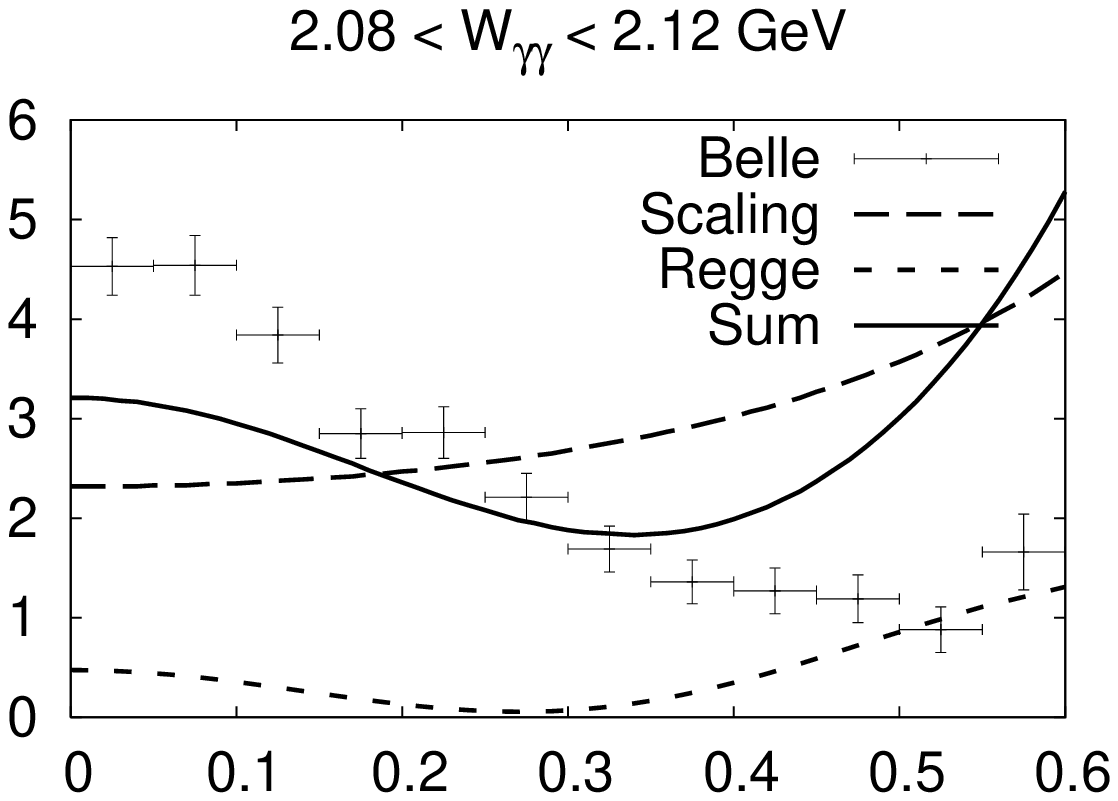,width=5cm}
 \epsfig{file=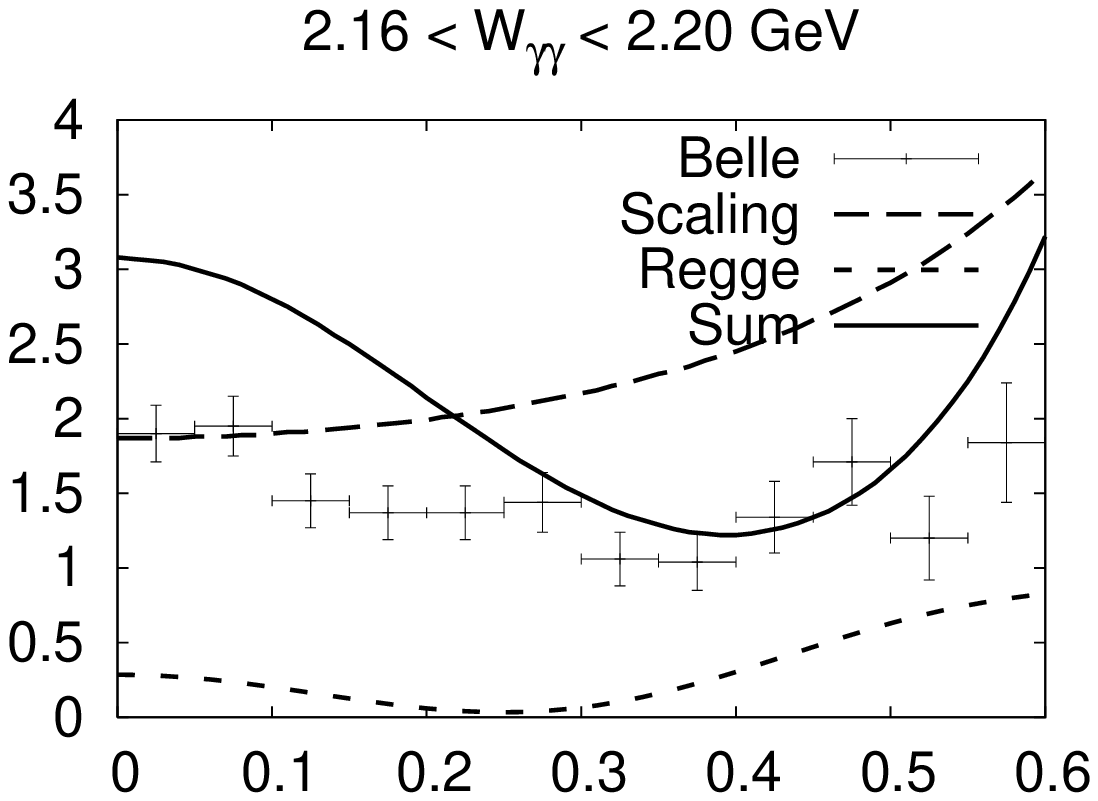,width=5cm}
 \end{center}
 \caption{The angular distribution of $\gamma\gamma\to K^+K^-$, at three 
representative energy ranges in between $1.40$ and $2.24$ GeV (upper 
row), and between $2.00$ and $2.20$~GeV (lower row).
  We show the Belle data against the addition of the scaling and Regge 
amplitudes.
  The vertical error-bars on the Belle data are statistical only.
 \label{fig_kpkm_lowreg_ad}}}\end{figure*}

  The plateau structure is weakened but still visible when we adopt the 
Regge limiting expression of the Veneziano amplitude, as shown in 
fig.~\ref{fig_kpkm_lowregge}.
  The situation with respect to the angular distribution improves in the 
Regge limiting case, especially above $2$~GeV. This is shown in 
fig.~\ref{fig_kpkm_lowreg_ad}.
  The improvement in the fit compared with the Veneziano amplitude is 
mostly due to the reduced size of the Regge limiting expression compared 
to the Veneziano amplitude. If the Regge amplitude normalization is 
modified to more closely resemble that of the Veneziano amplitude, the 
fit deteriorates.

  The angular distribution below $2$~GeV is not well reproduced.

  Since the Veneziano amplitude by itself fits the distribution well in 
the low-energy region, it is tempting to introduce a form factor that 
allows smooth transition between the long-distance and scaling 
amplitudes. We have experimented with several such possibilities and 
found that although the fit with the angular distribution improves, it 
is difficult to obtain the plateau structure of the integrated cross 
section.

  Let us now turn our attention to the $K_SK_S$ cross section. We make
two assumptions for the $K^0\overline{K^0}=K_SK_S+K_LK_L$ cross section
with respect to the $K^+K^-$ cross section, namely:
 \begin{enumerate}
  \item the scaling amplitude is scaled by a constant factor of
$1/4$;
  \item the Veneziano amplitude remains the same.
 \end{enumerate}
  The first implies the complete exclusion of the strange-quark 
contribution, and the second implies the invariance of the long-distance 
amplitude with respect to isospin.

\begin{figure}[ht]{
 \begin{center}
  $\sigma(\gamma\gamma\to KK)/$nb, $|\cos\theta^*|<0.6$\\
  \epsfig{file=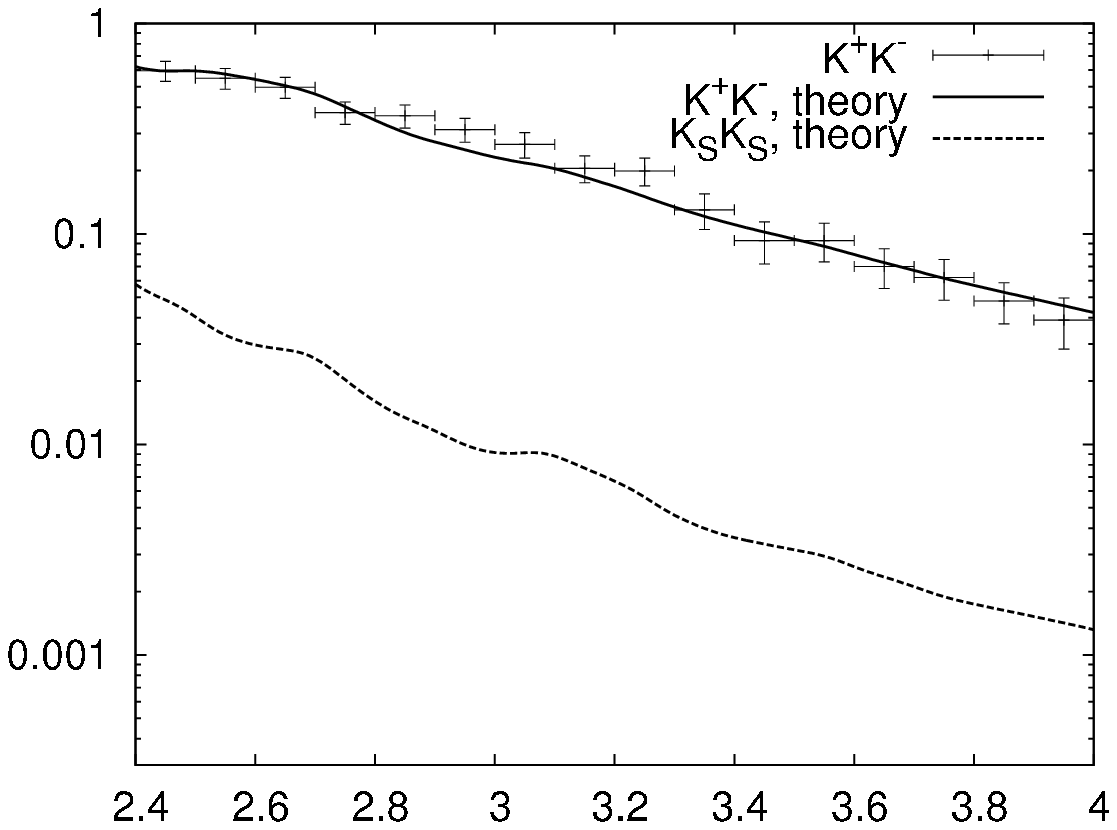,width=8cm}\\
  $W_{\gamma\gamma}/$GeV.
 \end{center}
 \caption{$\gamma\gamma\to KK$ integrated cross section in the region 
$|\cos\theta^*|<0.6$ , for $K^+K^-$ and $K_SK_S$.
  We show the region $2.4$~GeV$<W_{\gamma\gamma}<4$~GeV.
  We show the experimental results \cite{bellek} and the sum of the 
scaling and Veneziano model amplitudes.
 \label{fig_kpkm_ksks}}}\end{figure}

  The results are shown in fig.~\ref{fig_kpkm_ksks}. The large ratio 
between $K^+K^-$ and $K_SK_S$ cross sections, as well as the increasing 
ratio between the two, is reproduced. However, the ratio seems to 
increase more rapidly in the real data.
  From our above results of the $K^+K^-$ cross section, we expect that 
the angular distribution may not be reproduced correctly. However, if 
the long-distance effects are still active in the high-energy range, a 
model-independent statement is that the angular distribution would be 
affected, and will not be given by a simple scaling form such as 
eqn.~(\ref{eqn_scaling_angular_distribution}).

  We note that for pions, both in the analysis of 
sec.~\ref{sec_clebschgordan} and in the Regge/Veneziano amplitudes, 
isospin invariance is respected. Hence we expect:
 \begin{equation}
  \frac{\sigma(\gamma\gamma\to\pi^0\pi^0)}
  {\sigma(\gamma\gamma\to\pi^+\pi^-)}=\frac12,
 \end{equation}
  always, so long as the small difference in the neutral and charged 
pion masses can be neglected. The ratio would be violated by the 
inclusion of `cat's ears'-type diagrams.

 \section{Conclusions}

  We studied exclusive hadron pair photo-production processes in
low-energy photon--photon collision.

  Motivated by the experimental observation of the unexpectedly large
suppression of the ratio $K_SK_S/K^+K^-$, we looked into the $SU(3)$
structure of the couplings involved in these processes, adopting an
$s$-channel picture.

  We presented the calculations both for the nonet mesons and for the
octet baryons.
  We argued that the difference between $K_SK_S$ and $K^+K^-$ cross 
sections is due to the simultaneous presence of $f/f'$ and $a$ in the 
$s$-channel. The ratio is further enhanced when the $f'$ contribution is 
suppressed. We argued that this could be as large as $32$.


  We proceeded with a model in which the ratio applies predominantly to 
the part of the amplitude which obeys a scaling behaviour. We added to 
this a long-distance amplitude whose limiting behaviour is given by 
Regge theory. This latter amplitude tends to be invariant under isospin.

  For the $\gamma\gamma\to p\bar p$ process, adopting a Regge amplitude 
for the long-distance dynamics, we obtain a distribution that not only 
fits the integrated cross section well but reproduces the behaviour of 
the angular distribution.

  For $\gamma\gamma\to K^+K^-$, the long-distance dynamics was simulated 
using the Veneziano model. The summation of this and the scaling 
amplitude results in a curve for the integrated cross section that is 
similar to the experimental data. On the other hand, the angular 
distribution is not well reproduced. The Regge limiting amplitude has 
better fit with the data, mainly because of the reduced size of the 
amplitude compared to the Veneziano amplitude. In any case, the 
distribution near the $f'$ peak is reproduced better by the pure 
Veneziano/Regge amplitudes, possibly indicating the incompleteness in 
our parametrization of the resonance region.

 \acknowledgement

  Acknowledgements:
  We thank Chen-Cheng~Kuo for informative discussions on the
experimental results.
  RCV thanks Hai-Yang~Cheng for the kind invitation and support during
the stay at the Institute of Physics, Academia Sinica, where part of the 
work was done. RCV also gratefully acknowledges the financial support by 
the Physics Division of NCTS.
  KO thanks Augustine~Chen, Wan-Ting~Chen and Chun-Khiang~Chua for
discussions.

 \endacknowledgement


\begin{thebibliography}{88}

 \bibitem{bellep} C.-C.~Kuo {et al.} [Belle Collaboration], 
\plb{621}{2005}{41} [arXiv:\hepex{0503006}];\\
 C.-C.~Kuo, talk at Belle $\tau/2\gamma$ meeting, Nagoya, Japan, 11--12
March 2004.

 \bibitem{bellek} K.~Abe {et al.} [Belle Collaboration],
\epjc{32}{2003}{323} [arXiv:\hepex{0309077}];
H.~Nakazawa {et al.} [Belle Collaboration],
\plb{615}{2005}{39} [arXiv:\hepex{0412058}];

 \bibitem{qcount} V.A.~Matveev, R.M.~Muradian and A.N.~Tavkhelidze,
\ncl{7}{1973}{719};\\
S.J.~Brodsky and G.R.~Farrar,
\prd{11}{1975}{1309}.

 \bibitem{venus} H.~Hamasaki {et al.} [VENUS Collaboration],
\plb{407}{1997}{185}.

 \bibitem{cleo} M.~Artuso {et al.} [CLEO Collaboration],
\prd{50}{1994}{5484}.

 \bibitem{l3ppbar} P.~Achard {et al.} [L3 Collaboration],
\plb{571}{2003}{11} [arXiv:\hepex{0306017}].

 \bibitem{tpc} H.~Aihara {et al.} [TPC/Two-Gamma Collaboration],
\prl{57}{1986}{404}.

 \bibitem{argus} H.~Albrecht {et al.} [ARGUS Collaboration],
\zpc{48}{1990}{183}.

 \bibitem{belleks} W.T.~Chen {et al.} [Belle Collaboration] ,
arXiv:\hepex{0609042}.

 \bibitem{benayounchernyak}
  M.~Benayoun and V.L.~Chernyak, \npb{329}{1990}{285};
  V.L.~Chernyak, \plb{640}{2006}{246}.

 \bibitem{brodskylepage} S.J.~Brodsky and G.P.~Lepage,
\prd{24}{1981}{1808}.

 \bibitem{pinch} See ref.~\cite{brodskylepage} and the listing under 
ref.~5 therein.

 \bibitem{handbag}
  M.~Diehl, P.~Kroll and C.~Vogt,
\plb{532}{2002}{99} [arXiv:\hepph{0112274}];
\epjc{26}{2003}{567} [arXiv:\hepph{0206288}].

 \bibitem{collins} See, for example, P.D.B.~Collins, {An introduction 
to Regge theory and high energy physics}, Cambridge University Press, 
1977.

 \bibitem{ddln} A.~Donnachie, H.G.~Dosch, P.V.~Landshoff and O.~Nachtmann, 
{Pomeron physics and QCD}, Cambridge University Press, 2002.

 \bibitem{dl_tripleglu} A.~Donnachie and P.V.~Landshoff,
\npb{231}{1984}{189}.

 \bibitem{thetap} T.~Feldmann,
\ijmpa{15}{2000}{159} [arXiv:\hepph{9907491}].

 \bibitem{pdg} S.~Eidelman {et al.}  [Particle Data Group],
\plb{592}{2004}{1}.

 \bibitem{Nagels:1979xh} M.M.~Nagels {et al.},
\npb{147}{1979}{189};\\
S.~Kanwar, R.C.~Verma and M.P.~Khanna,
\ptp{62}{1979}{1152}.

 \bibitem{finite_energy_sumrule} K.~Igi, \prl{9}{1962}{76};\\
 R.~Dolen, D.~Horn and C.~Schmid, \prl{19}{1967}{402};\\
 K.~Igi and S.~Matsuda, \prl{18}{1967}{625};\\
 A.A.~Logunov, L.D.~Soloviev and A.N.~Tavkhelidze, \prl{24B}{1967}{181}.

 \bibitem{odagiriveneziano} K.~Odagiri,
\npa{748}{2005}{168} [arXiv:\hepph{0406267}].

 \bibitem{veneziano} G.~Veneziano,
\nc{57}{1968}{190}.

 \bibitem{storrow} J.~K.~Storrow, 
\prep{103}{1984}{317}.

 \bibitem{baryon_l3} P.~Achard {et al.} [L3 Collaboration],
\plb{536}{2002}{24} [arXiv:\hepex{0204025}].

 \bibitem{baryon_cleo} S.~Anderson {et al.} [CLEO Collaboration],
\prd{56}{1997}{2485} [arXiv:\hepex{9701013}].

 \bibitem{berger-schweiger} C.F.~Berger and W.~Schweiger,
\epjc{28}{2003}{249} [arXiv:\hepph{0212066}].

 \end{thebibliography}
 \end{document}